\documentclass{article}
\def\ftoday{{\sl  \number\day \space\ifcase\month
\or Janvier\or F\'evrier\or Mars\or avril\or Mai
\or Juin\or Juillet\or Ao\^ut\or Septembre\or Octobre
\or Novembre \or D\'ecembre\fi
\space  \number\year}}
\newcommand{\journal}[4]{{\em #1~}#2\,(19#3)\,#4;}

\newcommand{\hpa}{\journal {Helv. Phys. Acta}}

\newcommand{\ijmp}{\journal {Int. J. Mod. Phys.}}

\newcommand{\pr}{\journal {Phys. Rev.}}

\newcommand{\prl}{\journal {Phys. Rev. Lett.}}

\newcommand{\cmp}{\journal {Comm. Math. Phys.}}
\newcommand{\cqg}{\journal {Class. Quantum Grav.}}

\newcommand{\np}{\journal {Nucl. Phys.}}
\newcommand{\pl}{\journal {Phys. Lett.}}
\newcommand{\mpl}{\journal {Mod. Phys. Lett.}}
\newcommand{\prep}{\journal {Phys. Reports}}

\newcommand{\annp}{\journal {Ann. Phys. (N.Y.)}}

\def\appendix{\par
 \setcounter{chapter}{0}
 \setcounter{section}{0}
 \def\@chapapp{Appendice}                                 % Appendice
 \def\thechapter{\Alph{chapter}}}
\def\tableofcontents{\@restonecolfalse\if@twocolumn\@restonecoltrue
 \onecolumn\fi\chapter*{Table des mati\`eres%
 \@mkboth{TABLE DES MATIERES}{TABLE DES MATIERES}}
 \@starttoc{toc}\if@restonecol\twocolumn\fi}              % Table des mati\`eres
\makeatletter
\@addtoreset{equation}{section}
\makeatother
\renewcommand{\theequation}{\thesection.\arabic{equation}}

\newcommand{\es}{\\[3mm]}
\newcommand{\eq}{\begin{equation}}
\newcommand{\eqn}[1]{\label{#1}\end{equation}}
\newcommand{\eea}{\end{eqnarray}}
\newcommand{\eqap}{\begin{eqnarray}}
\newcommand{\eqanp}[1]{\label{#1}\end{eqnarray}}

\newcommand{\eqan}[1]{\ea\label{#1}\end{equation}}
\newcommand{\equ}[1]{(\ref{#1})}
\newcommand{\ba}{\begin{array}}
\newcommand{\ea}{\end{array}}
\newcommand{\eqac}{\begin{equation}\begin{array}{rcl}}
\newcommand{\eqacn}[1]{\end{array}\label{#1}\end{equation}}

\newcommand{\dis}{\displaystyle}

\def\LP{\displaystyle{\Biggl(}}
\def\RP{\displaystyle{\Biggr)}}
\newcommand{\lp}{\left(}\newcommand{\rp}{\right)}
\newcommand{\lc}{\left[}\newcommand{\rc}{\right]}
\newcommand{\lac}{\left\{}\newcommand{\rac}{\right\}}
\newcommand{\Dleft}[1]{
{\raise .7ex \hbox{${\raise -.7ex \hbox{$#1$}}^{^{\kern-7pt \leftarrow}}$}}
}

\newcommand{\chib}{{{\bar\chi}}}
\newcommand{\G}{\Gamma}
\newcommand{\D}{\Delta}
\renewcommand{\a}{\alpha}
\renewcommand{\b}{\beta}
\renewcommand{\d}{\delta}

\newcommand{\e}{\varepsilon}
\newcommand{\eb}{{\bar\varepsilon}}
\newcommand{\f}{\phi}
\newcommand{\F}{\Phi}

\newcommand{\g}{\gamma}
\renewcommand{\k}{\kappa}
\renewcommand{\l}{\lambda}
\newcommand{\lb}{{\bar\lambda}}
\newcommand{\Lb}{{\bar\Lambda}}
\renewcommand{\L}{\Lambda}
\newcommand{\m}{\mu}
\newcommand{\n}{\nu}

\newcommand{\om}{\omega}
\renewcommand{\O}{\Omega}

\newcommand{\p}{\psi}
\newcommand{\pb}{{\bar\psi}}
\renewcommand{\r}{\rho}
\newcommand{\s}{\sigma}
\renewcommand{\S}{\Sigma}

\newcommand{\BB}{{\cal B}}
\newcommand{\BS}{{\cal B}_\Sigma}

\newcommand{\FF}{{\cal F}}
\newcommand{\GG}{{\cal G}}
\newcommand{\HH}{{\cal H}}
\newcommand{\II}{{\cal I}}

\renewcommand{\SS}{{\cal S}}

\newcommand{\cb}{{\bar c}}
\newcommand{\fb}{{\bar\f}}

\newcommand{\da}{{\dot{\a}}}
\newcommand{\db}{{\dot{\b}}}
\newcommand{\dg}{{\dot{\g}}}

\newcommand{\vf}{{\varphi}}

\newcommand{\complex}{{\kern .1em {\raise .47ex
                       \hbox {$\scriptscriptstyle |$}}
                       \kern -.4em {\rm C}}}
\newcommand{\real}{{{\rm I} \kern -.19em {\rm R}}}
\newcommand{\rational}{{\kern .1em {\raise .47ex
                        \hbox{$\scripscriptstyle |$}}
                        \kern -.35em {\rm Q}}}
\renewcommand{\natural}{{\vrule height 1.6ex width
                         .05em depth 0ex \kern -.35em {\rm N}}}
\newcommand{\znatural}{{ {\rm Z}\kern -.34em {\rm Z}}}

\def\intx{\displaystyle{\int d^4 \! x \, }}

\newcommand{\half}{\frac 1 2}

\newcommand{\pa}{\partial}

\newcommand{\ddsum}[2]{\dis{\sum_{#1}^{#2}}}
\newcommand{\dsum}[1]{\dis{\sum_{#1}}}
\newcommand{\dpad}[2]{{\displaystyle{\frac{\pa #1}{\pa #2}}}}
\newcommand{\dfud}[2]{{\displaystyle{\frac{\delta #1}{\delta #2}}}}
\newcommand{\fud}[2]  {{\displaystyle{\frac{\delta #1}{\delta #2}}}}
\newcommand{\dfrac}[2]{{\displaystyle{\frac{#1}{#2}}}}

\newcommand{\twiddle}{\lower.9ex\rlap{$\kern -.1em\scriptstyle\sim$}}

\def\un{1\kern-3pt \rm I}
\def\ddouble{{\pa^{^{\kern-6pt \leftrightarrow}}}}
\newcommand{\bstar}{\begin{petit} \noindent {\Large $\star$} }
\newcommand{\estar}{\end{petit}}

\newcommand{\ie}{{{\em i.e.}\ }}

\newcommand{\BSn}[1]{\BS^{(#1)}}

\def\B0modd{\makebox[28mm]{$\BSn{0}$-modulo-$d$}}

\newcommand{\sla}{\raise.15ex\hbox{$/$}\kern -.57em}
\newcommand{\Sla}{\raise.15ex\hbox{$/$}\kern -.70em D}

\newcommand{\sbar}{{\bar \s}}
\newcommand{\lx}{{\cal L}_\xi}
\newcommand{\vb}[2]{{\rm e}_{#1}{}^{#2}}
\newcommand{\vbb}[2]{{\rm e}_{#1}{}_{#2}}
\newcommand{\vbh}[2]{{\rm e}^{#1}{}^{#2}}
\newcommand{\gP}{\Psi}
\newcommand{\gPb}{{\bar \Psi}}

\def\ksl{{\kern.12em {\raise.3ex\hbox{/} \kern-.59em k}}}
\def\psl{{\kern.12em {\raise.3ex\hbox{/} \kern-.59em p}}}
\def\dsl{{\kern.1em {\raise.3ex\hbox{/} \kern-.58em \partial}}}
\def\Asl{{\kern.12em {\raise.3ex\hbox{/} \kern-.77em A}}}
\def\Dsl{{\kern.15em {\raise.3ex\hbox{/} \kern-.7em D}}}

\newcommand{\kzero}{{\k\kern-.45em \raisebox{2mm}{\tiny o}\kern.30em}}
\newcommand{\lkzero}{{\k\kern-.35em \raisebox{1.6mm}{\tiny o}\kern.30em}}
\def\@@begthe#1{\@ifnextchar[{\@optbegthe#1}{\@begthe#1}}
\def\@begthe#1{ #1}
\def\@optbegthe#1[#2]{ {#2} #1}
\newcommand{\newthe}[6]{
 \def\nlni{\par\ifvmode\removelastskip\fi\vskip\baselineskip\noindent}
 \def\xxxend{\endgroup\vskip\baselineskip}
 \newenvironment{#1}{\nlni\begingroup\refstepcounter{#4}#5#3
                     \thechapter.\arabic{#4}\@@begthe{#6}}{\xxxend}
 \newenvironment{#2}{\nlni\begingroup#5#3\@@begthe{#6}}{\xxxend}}

\newthe{lemme}{lemme*}{Lemme}{lemmacount}{\bf}{\it}
\newthe{proposition}{proposition*}{Proposition}{lemmacount}{\bf}{\it}
\newthe{theoreme}{theoreme*}{Th\'eor\`eme}{lemmacount}{\bf}{\it}
\newthe{corollaire}{corollaire*}{Corollaire}{lemmacount}{\bf}{\it}

\newthe{exercice}{exercice*}{Exercice}{lemmacount}{\bf}{\it}
\newthe{probleme}{probleme*}{Probl\`eme}{lemmacount}{\bf}{\it}
\newthe{solution}{solution*}{Solution}{lemmacount}{\bf}{\it}
\newthe{definition}{definition*}{Definition}{lemmacount}{\bf}{\it}

\newthe{exemple}{exemple*}{Exemple}{lemmacount}{\bf}{\rm}
\newthe{note}{note*}{Note}{lemmacount}{\bf}{\rm}
\newthe{question}{question*}{Question}{lemmacount}{\bf}{\rm}
\def\bbbc{{\mathchoice {\setbox0=\hbox{$\displaystyle\rm C$}\hbox{\hbox
to0pt{\kern0.4\wd0\vrule height0.9\ht0\hss}\box0}}
{\setbox0=\hbox{$\textstyle\rm C$}\hbox{\hbox
to0pt{\kern0.4\wd0\vrule height0.9\ht0\hss}\box0}}
{\setbox0=\hbox{$\scriptstyle\rm C$}\hbox{\hbox
to0pt{\kern0.4\wd0\vrule height0.9\ht0\hss}\box0}}
{\setbox0=\hbox{$\scriptscriptstyle\rm C$}\hbox{\hbox
to0pt{\kern0.4\wd0\vrule height0.9\ht0\hss}\box0}}}}
\def\bbbe{{\mathchoice {\setbox0=\hbox{\smalletextfont e}\hbox{\raise
0.1\ht0\hbox to0pt{\kern0.4\wd0\vrule width0.3pt
height0.7\ht0\hss}\box0}}
{\setbox0=\hbox{\smalletextfont e}\hbox{\raise
0.1\ht0\hbox to0pt{\kern0.4\wd0\vrule width0.3pt
height0.7\ht0\hss}\box0}}
{\setbox0=\hbox{\smallescriptfont e}\hbox{\raise
0.1\ht0\hbox to0pt{\kern0.5\wd0\vrule width0.2pt
height0.7\ht0\hss}\box0}}
{\setbox0=\hbox{\smallescriptscriptfont e}\hbox{\raise
0.1\ht0\hbox to0pt{\kern0.4\wd0\vrule width0.2pt
height0.7\ht0\hss}\box0}}}}
\def\bbbq{{\mathchoice {\setbox0=\hbox{$\displaystyle\rm Q$}\hbox{\raise
0.15\ht0\hbox to0pt{\kern0.4\wd0\vrule height0.8\ht0\hss}\box0}}
{\setbox0=\hbox{$\textstyle\rm Q$}\hbox{\raise
0.15\ht0\hbox to0pt{\kern0.4\wd0\vrule height0.8\ht0\hss}\box0}}
{\setbox0=\hbox{$\scriptstyle\rm Q$}\hbox{\raise
0.15\ht0\hbox to0pt{\kern0.4\wd0\vrule height0.7\ht0\hss}\box0}}
{\setbox0=\hbox{$\scriptscriptstyle\rm Q$}\hbox{\raise
0.15\ht0\hbox to0pt{\kern0.4\wd0\vrule height0.7\ht0\hss}\box0}}}}
\def\bbbt{{\mathchoice {\setbox0=\hbox{$\displaystyle\rm
T$}\hbox{\hbox to0pt{\kern0.3\wd0\vrule height0.9\ht0\hss}\box0}}
{\setbox0=\hbox{$\textstyle\rm T$}\hbox{\hbox
to0pt{\kern0.3\wd0\vrule height0.9\ht0\hss}\box0}}
{\setbox0=\hbox{$\scriptstyle\rm T$}\hbox{\hbox
to0pt{\kern0.3\wd0\vrule height0.9\ht0\hss}\box0}}
{\setbox0=\hbox{$\scriptscriptstyle\rm T$}\hbox{\hbox
to0pt{\kern0.3\wd0\vrule height0.9\ht0\hss}\box0}}}}
\def\bbbs{{\mathchoice
{\setbox0=\hbox{$\displaystyle     \rm S$}\hbox{\raise0.5\ht0\hbox
to0pt{\kern0.35\wd0\vrule height0.45\ht0\hss}\hbox
to0pt{\kern0.55\wd0\vrule height0.5\ht0\hss}\box0}}
{\setbox0=\hbox{$\textstyle        \rm S$}\hbox{\raise0.5\ht0\hbox
to0pt{\kern0.35\wd0\vrule height0.45\ht0\hss}\hbox
to0pt{\kern0.55\wd0\vrule height0.5\ht0\hss}\box0}}
{\setbox0=\hbox{$\scriptstyle      \rm S$}\hbox{\raise0.5\ht0\hbox
to0pt{\kern0.35\wd0\vrule height0.45\ht0\hss}\raise0.05\ht0\hbox
to0pt{\kern0.5\wd0\vrule height0.45\ht0\hss}\box0}}
{\setbox0=\hbox{$\scriptscriptstyle\rm S$}\hbox{\raise0.5\ht0\hbox
to0pt{\kern0.4\wd0\vrule height0.45\ht0\hss}\raise0.05\ht0\hbox
to0pt{\kern0.55\wd0\vrule height0.45\ht0\hss}\box0}}}}
\newcommand{\nmGA}{{\G\kern-.45em \raisebox{3.3mm}{\tiny o}\kern.30em}}
\newcommand{\nmSI}{{\S\kern -.50em \raisebox{3.3mm}{\tiny o}\kern.30em}}
\newcommand{\nmSS}{{\SS\kern -.50em \raisebox{3.3mm}{\tiny o}\kern.30em}}
\newcommand{\nmBB}{{\BB\kern -.50em \raisebox{3.3mm}{\tiny o}}}
\newcommand{\nmGAind}{{\G\kern-.37em\raisebox{2.1mm}{\tiny o}}}
\newcommand{\nmSIind}{{\S\kern-.37em\raisebox{2.1mm}{\tiny o}}}
\newcommand{\nnmGAind}{{\kern.15em\nmGAind}}
\newcommand{\nnmSIind}{{\kern.10em\nmSIind}}
\newcommand{\nmBnmG}{{\nmBB_\nnmGAind\kern.20em}}
\newcommand{\nmBnmS}{{\nmBB_\nnmSIind\kern.20em}}
\newcommand{\nmBS}{{\nmBB_{\kern.5mm\S}}}
\newcommand{\nmBG}{{\nmBB_{\kern.5mm\G}}}
\newcommand{\nmBF}{{\nmBB_{\kern.5mmF}}}

\newcommand{\Bv}{\displaystyle{\biggl\vert}}
\newcommand{\Lac}{\displaystyle{\biggr\{}}
\newcommand{\Rac}{\displaystyle{\biggr\}}}
\newcommand{\LAC}{\displaystyle{\Biggl\{}}
\newcommand{\RAC}{\displaystyle{\Biggr\}}}
\newcommand{\fl}[1]{{\lp{#1}\rp_{\mbox{\tiny{F.L}}}}}
\newcommand{\point}[1]{\vspace{3mm}
\noindent{\bf #1}}
\newcommand{\wtd}{{\widetilde d}}
\newcommand{\sumall}{\displaystyle{\sum_{\mbox{\tiny{all fields}}}}}
\newcommand{\hepth}[1]{hep-th/\,#1}
\newcommand{\hepph}[1]{hep-ph/\,#1}
\renewcommand{\journal}[4]{{\em #1~}#2\,(19#3)\,#4}
\newcommand{\cohom}[2]{\HH(#1,#2)}
\newcommand{\cohomext}[2]{\HH_{\rm ext}(#1,#2)}

\newcommand{\invbs}[2]{\II(#1,#2)}

\begin{document}
\parindent=0cm
\noindent

%sugra.tex\hfill\ftoday

%%%%%%%%%%%%%%%%%%%%%%%%%%%%%%%%%%%%%%%%%%%%%%%%
{\hfill\parbox{50mm}{\large hep-th/9802027\\
                            UFES--DF--OP98/1\\
                            UGVA-DPT 1998/01-997}} \vspace{3mm}

\begin{center}

{\huge{\bf The Supercurrent Trace Identities of the \es
           $N=1$, $D=4$ Super-Yang-Mills Theory \es
             in the Wess-Zumino Gauge}}

\vspace{1cm}

{\Large Olivier Piguet\footnote{Supported by
  the {\it Conselho Nacional de
  Desenvolvimento Cient\'\i fico e Tecnol\'ogico (CNPq), Brazil.}
  \\ In leave of absence from the
  {\it D\'epartement de Physique Th\'eorique --
       Universit\'e de Gen\`eve, Switzerland.}}$^{,2}$}
\vspace{3mm}

{\it Universidade Federal do Esp\'\i rito
Santo (UFES),
Departamento de F\'\i sica, \\Campus Universit\'ario de Goiabeiras -
29060-900 - Vit\'oria - ES - Brazil}
\vspace{5mm}

{\Large and Sylvain Wolf\footnote{Supported in part by the
{\it Swiss National Science Foundation.}}}
\vspace{3mm}

{\it D\'epartement de Physique Th\'eorique --
     Universit\'e de Gen\`eve\\24, quai E. Ansermet -- CH-1211 Gen\`eve
     4 -- Switzerland}

\end{center}
\vspace{10mm}

\begin{center}
{\large{\bf Abstract}}
\end{center}

{\it The supercurrent components of the $N=1$, $D=4$  Super-Yang-Mills
theory in the Wess-Zumino gauge
are coupled to the components of a background supergravitation
field in the ``new minimal'' representation, in order to describe the
various conservation laws in a functional way through the Ward
identities for the diffeomorphisms and for the local supersymmetry,
Lorentz and $R$-transformations. We also incorporate in the
same functional formalism the supertrace identities, which
leads however to a slight modification
of the new minimal representation for supergravity, thus leading
to a conformal version of it.
The most general classical action obeying all the symmetry
constraints and the condition of power-counting renormalizability
is constructed.
}
%\vfill

%\newpage

%%%%%%%%%%%%%%%%%%%%%%%%%%%%%%%%%%%%%%%%%%%%%%%
\section{Introduction}
 The renormalization properties of the
supercurrent~\cite{fer-zum,ps-book} in
$N=1$ supersymmetric gauge theories (Super Yang-Mills -- SYM) are of
fundamental importance for the characterization of
the ultraviolet behavior of the latter.
Indeed, the supercurrent multiplet contains, together with the conserved
spinor current associated to supersymmetry, the axial current associated
to $R$-invariance and the {\it improved} conserved
energy-momentum tensor. This implies that the anomaly of the axial current
and the trace anomaly of the energy-momentum tensor -- or scale anomaly,
which controls the ultraviolet asymptotic behavior of the
theory~\cite{as-freedom} --
are  related by supersymmetry. These anomalies appear as
breakings of a
set of ``supertrace identities'' including the usual energy-momentum
trace identity, the spinor current trace identity and the divergence of
the $R$ axial current~\cite{fer-zum}.

Thus, the use of the nonrenormalization theorem for the axial
current~\cite{a-b-theorem,lps:ab-theorem,pi-so} leads, through
supersymmetry,
to strong constraints on the scale anomaly, and
even, in certain cases~\cite{lps:beta=0}, to its
absence\footnote{See \cite{so-we(n=4)} for an earlier application of
this idea to a formal proof of the
finiteness of $N=4$ super-Yang-Mills.}.

The references~\cite{lps:beta=0} deal with this
problem in the superfield
formalism, which is well adapted if supersymmetry is
exact\footnote{See \cite{kazakov} for a superfield approach,
not relying
on the supercurrent anomalies,
and with application to  broken supersymmetry.}.
 But it is
highly desirable to extend the discussion to the more realistic
situation where supersymmetry is broken, possibly by mass terms. There,
most of the advantages of superspace being lost, it is worthwhile to
work in the component formalism, and better in the Wess-Zumino gauge
where the number of component fields is minimal.
As a preliminary step to this program, the renormalizability
of the SYM theories in the Wess-Zumino gauge has been recently
established in~\cite{white1,white2,I} for the unbroken case\footnote{See
\cite{white2,magg1,magg2} for extended supersymmetry
$N=2$ and $N=4$.}  and
in~\cite{II} in the presence of ``soft'' breakings by mass terms.

The aim of the present paper is
to set up a functional formalism
describing, through Ward identities to be obeyed by the generating
functional of the Green functions, both, the conservation laws and
the supertrace identities  for the components of the supercurrent.
We shall do it in the component formalism, in the Wess-Zumino gauge,
proceeding in two steps.

The first step of our approach consists in the coupling of the supercurrent
components to the components of a supermultiplet of supergravity
fields considered as external (background) fields. The use of the
so called ``new minimal'' realization of supergravity~\cite{sohn-west}
allows to express all the conservation laws as the Ward identities
for the local symmetries -- diffeomorphisms, Lorentz, supersymmetry, $R$
-- of this model, which now is one of gauge and matter fields coupled
to a classical
supergravity background. We shall begin by  reestablishing
the associated BRS formulation~\cite{brs-gauge}
 of these local symmetries~\cite{bau-bel} in our
proper notations, taking into account the
presence of the gauge and matter fields.
We shall then check the off-shell nilpotency of the
BRS operator after the introduction of the external fields coupled to
those BRS transformations which are nonlinear in the dynamical fields
- the so-called ``Batalin-Vilkoviski antifields''~\cite{bv} --
and write down the Slavnov-Taylor
identity, a nonlinear Ward identity containing
all the information about
the symmetries and their algebra, hence on the supercurrent conservation
laws and the supercurrent algebra. The achievement of this program
represents the supersymmetric generalization of the results
found in~\cite{kraus-sib} for the algebra of the
energy-momentum tensor alone.

The second step is the incorporation of the supertrace identities into
the formalism. As we shall see, in order to avoid hard
breakings of some of the identities,
we have to proceed to a slight
reformulation of the coupling to the external supergravity fields,
leading to a modified realization of the supergravity
algebra with new symmetries
corresponding to the supertrace identities.
 This actually represents a particular formulation
of conformal supergravity coupled to
matter\footnote{As far the authors know, this particular formulation of
conformal supergravity did not yet appear in the literature.}
We then give a
full  BRS formulation of this new algebra, with an off-shell nilpotent BRS
operator and a new Slavnov-Taylor identity.

We construct the most general --  power-counting renormalizable --
action
obeying the Slavnov-Taylor identity, and find in particular that the
pure supergravitational part is the known action of conformal
supergravity~\cite{csg,frad-tsey}.

We finally check the stability of the theory under small perturbations as a
preliminary step towards the proof of the renormalizability~\cite{pi-wo}.
Although the present paper deals exclusively with the
classical theory, the stability proof needs only to be complemented by
the study of the possible anomalies in order to
prove the renormalizability of the theory and to characterize its
ultraviolet properties.

The plan of the paper is as follows. In Section 2 we introduce the coupling of
matter and Yang-Mills supermultiplets with the supergravity multiplet
in the ``new minimal'' formulation. (Super)trace identities and Weyl
symmetry are discussed together with their breakings in Section 3
within the same formalism, whereas
the new formulation is introduced in Section 4 where the
full new BRS algebra is given, too. Section 5 is devoted to the derivation
of the most general action, including its pure conformal supergravity
part, and to the proof of the
stability of the theory under small perturbations.

Some technicalities are contained in the Appendices A to G.

%%%%%%%%%%%%%%%%%%%%%%%%%%%%%%%%%%%%%%%%%%%%%%%%%%%%%
\section{$N=1$ Local SYM on Curved Space-time}\label{sym-local}

As explained in the Introduction, we want to extend the previously
studied model,
$N=1$ massless (global)  SYM in the
Wess-Zumino gauge~\cite{I}, to the context of
field theory on curved space-time.
This means that we want to construct a model which is invariant
under the {\em local} superpoincar\'e group rather than under the
{\em global} superpoincar\'e group.
But doing this is nothing but dealing with Supergravity
coupled to SYM, with the crucial
difference that we will consider the vierbein and the gravitino fields
as
{\em  external} fields and doing so, we will not face the problem of
dealing with a model which is power-counting non-renormalizable.
In this way, we see that the title of this section could also have been
``External $N=1$ Supergravity coupled to SYM".

\subsection{Field Content of New Minimal $N=1$ Supergravity}

As one of the goal of this paper is to exhibit the link between the
conservation law of the $R$-current and the trace of the energy-momentum
tensor via the supertrace identity, we choose the {\em new} minimal
formulation of supergravity~\cite{sohn-west} since it contains the
source of
the $R$-current, namely the field $B_\m$,  as
a gauge field. The vierbein and gravitino fields will be the sources
of the energy-momentum tensor and of the supersymmetry current,
respectively.

The field content of this formulation is~\cite{sohn-west}:
\begin{enumerate}
\item[-] the vierbein\footnote{ See \ref{appendixb}.} $\vb{\m}{a}$,
\item[-] the gravitino $\gP_{\m\a}$,
\item[-] the gauge field $B_\m$ of the $R$-gauge transformations,
\item[-] the gauge field $C_{\m\n}$ of the $C$-gauge transformations.
\end{enumerate}

The ghost numbers, Grassmann parities, dimensions and $R$-weights
of  all the fields appearing in this paper
are given in \ref{appendicetableau}.  Our notations and
conventions are given in \ref{appendixa}.

\underline{Remark}: This is a torsion free formulation of gravity,
which means that the Lorentz connection $w_{\m ab}$ is a function
of the vierbein (see \ref{appendixb}
for a summary of the vierbein formalism).

The generalized BRS
transformations~\cite{I,dixon,white1,white2,magg1,magg2,mpr},
including diffeomorphisms, local
Lorentz transformations, local supersymmetry and local $R$- and
$C$-transformations are given by~\cite{sohn-west}

\eq\ba{lll}
s \vb{\m}{a}&=& \lx\vb{\m}{a}
                 +\l^a {}_b\vb{\m}{b}
                 %-\kzero\vb{\m}{a}
                 +2\e\s^a\gPb_\m
                 +2\gP_\m\s^a\eb\es
s \gP_\m&=& \lx\gP_\m
             +\frac{i}{4}\l_{ab}\gP_\m\s^{ab}
             -i\eta \gP_\m
             %-\frac{1}{2}\kzero\gP_\m
             -i{\cal D}_\m \e\es
s \gPb_\m&=& \lx\gPb_\m
              -\frac{i}{4}\l_{ab}{\bar\s}^{ab}\gPb_\m
              +i\eta \gPb_\m
              %-\frac{1}{2}\kzero\gPb_\m
              -i{\cal D}_\m \eb\es
s  C_{\m\n}&=& \lx C_{\m\n}
               +\pa_\m w_\n
               -\pa_\n w_\m
               %-2\kzero C_{\m\n}
               \es &&
               -2\e\s_\m\gPb_\n
               -2\gP_\n\s_\m\eb
               +2\e\s_\n\gPb_\m
               +2\gP_\m\s_\n\eb  \es
s  B_\m&=& \lx B_\m
           +\pa_\m \eta
           -\e\s_\m\sbar^{\n\r}{\cal D}_\n\gPb_\r
           -{\cal D}_\n\gP_\r\s^{\n\r}\s_\m\eb \ ,
\ea\eqn{brscomplet2}

where $\xi^\m , \l_{ab} , \e , \eta$ and $w_\m$ are the ghosts for the
diffeomorphisms, Lorentz transformations, supersymmetry,
$R$- and $C$-transformations respectively -- i.e., their infinitesimal
parameters, but with opposite statistics.
The $\s^\m$ are the Pauli matrices and $\lx$ is the Lie derivative.

Moreover, the following condensed notations have been used

\eq\ba{lll}
{\cal D}_\m\e &=& \nabla_\m\e
                  +\frac{i}{4}F_{\m ab} \e\s^{ab}
                  +\frac{i}{4}V_\m\e
                  +\frac{1}{8}V^\l\e\s_{\m\l} \es
{\cal D}_\m\eb &=& \nabla_\m\eb
                  -\frac{i}{4}F_{\m ab} \sbar^{ab}\eb
                  -\frac{i}{4}V_\m\eb
                  +\frac{1}{8}V^\l\sbar_{\m\l}\eb \es
{\cal D}_\m\gP_\n &=& \nabla_\m\gP_\n
                  +\frac{i}{4}F_{\m ab} \gP_\n\s^{ab}
                  +\frac{i}{4}V_\m\gP_\n
                  +\frac{1}{8}V^\l\gP_\n\s_{\m\l} \es
{\cal D}_\m\gPb_\n &=& \nabla_\m\gPb_\n
                  -\frac{i}{4}F_{\m ab} \sbar^{ab}\gPb_\n
                  -\frac{i}{4}V_\m\gPb_\n
                  +\frac{1}{8}V^\l\sbar_{\m\l}\gPb_\n \es
F_{\m ab}&=&i\lp
     \gP_a\s_b\gPb_\m
   -  \gP_b\s_a\gPb_\m
   +  \gP_a\s_\m\gPb_b
   -  \gP_b\s_\m\gPb_a
   +  \gP_\m\s_a\gPb_b
   -  \gP_\m\s_b\gPb_a   \rp\es
V^\m &=& X^\m + 4Y^\m\es
X^\m &=&\e^{\m\n\r\l}\pa_\n C_{\r\l} \es
Y^\m &=&i\e^{\m\n\r\l} \gP_\n\s_\r\gPb_\l\ ,
\ea\eqn{dercovsupersym}

where the symbol $\nabla_\m$ means covariant derivative
with respect to usual gauge transformations (see subsection
\ref{dynfields}),
diffeomorphisms, Lorentz transformations and $R$-transformations
(see \ref{appendixc} for an explicit form of the covariant derivative).

Finally, the local algebra satisfied by these
transformations~\cite{sohn-west}
is encoded in the BRS transformations of the ghosts~\cite{bau-bel}:

\eq\ba{lll}
s \xi^\m&=&  \xi^\l \pa_\l \xi^\m
             +2iE^\m\es
s \l_{ab}&=& \lx\l_{ab}
             +(\l^2)_{ab}
             +2iE^\l {\widetilde w}_{\l ab}
             -\frac{i}{2} \e_{ab\m\n} E^\m V^\n\es
s \e&=& \lx \e
        +\frac{i}{4}\l_{ab}\e\s^{ab}
        -i\eta \e
        %-\frac{1}{2}\kzero\e
        +2E^\l\gP_\l\es
s \eb&=& \lx \eb
         -\frac{i}{4}\l_{ab}{\bar\s}^{ab}\eb
         +i\eta \eb
         %-\frac{1}{2}\kzero\eb
         +2E^\l\gPb_\l  \es
s \eta&=& \lx \eta
          -2iE^\l B_\l\es
s  w_\m&=& \lx w_\m
           +\pa_\m w
           %-2\kzero w_\m
           +2i E_\m
           +2iC_{\m\n}E^\n\es
s  w&=& \lx w
        %-2\kzero w
        -2iE^\l w_\l \ ,
\ea\eqn{brscomplet3}

where $w$ is the ghost for the ghost $w_\m$ and

\eq\ba{lll}
E^\m &=& \e\s^\m \eb\es
{\widetilde w}_{\m ab} &=& w_{\m ab} + F_{\m ab}\ .
\ea\eqn{emuetwtilde}

Thanks to these transformation laws for the ghosts,
this BRS operator $s$ is nilpotent:

\eq
s^2 = 0\ .
\eqn{nilpotencedes}

\underline{Remark}: Let us note the presence in the ghost
transformations of
gauge field dependent terms, like $s\eta = \cdots -2i E^\m B_\m$,
which means that
the anticommutator of two supersymmetries gives field dependent
gauge transformations. This accounts for the fact that
the  supersymmetry algebra in the ordinary formulation --
i.e. without ghosts -- is not closed but {\em infinite}.
This fact, already encountered in the rigid model~\cite{I}, is typical
of the Wess-Zumino gauge~\cite{wzgauge}.

\subsection{Field Content of Super Yang-Mills and Matter}
\label{dynfields}

The field content of the dynamical sector is~\cite{I,II}:

\begin{enumerate}
\item[-] the gauge and gaugino fields $A^i_\m$ and $\l^i$,
in the adjoint representation of the gauge group $\GG$,
which is supposed to be simple,
\item[-] the scalar and fermion matter fields $\f_A$ and $\p_A$, in some
(reducible) representation of $\GG$ carried by the
anti-Hermitian matrices $T^i_{AB}$,
\item[-] the ghost, antighost and Lagrange multiplier
fields $c^i$, $\cb^i$ and $b^i$, in the adjoint representation.
\end{enumerate}

 The BRS  transformations of these fields, including as well
the usual BRS  transformations associated with $\GG$ (called
$\GG$-gauge transformations in this paper),
are\footnote{The action of $s$ on the gravitational fields
are not modified by the adjunction of the usual BRS transformations
associated with $\GG$ since these
fields are $\GG$-gauge invariant.}

\eq\ba{lll}
s  A^{i}_\m &=& \lx A_\m^i
                + \nabla_\m c^i
                +\e\s_\m\lb^i
                +\l^i\s_\m\eb\es
s  \l^i   &=& \lx\l^i
               -f^{ijk} c^j \l^k
               +\frac{i}{4}\l_{ab}\l^i\s^{ab}
               -i\eta\l^i
               %+\frac{3}{2}\kzero\l^i
               -\frac{1}{2}\e\s^{\m\n}{\widetilde F}^i_{\m\n}
               -\frac{i}{2}G^2(\fb_AT^i_{AB}\f_B)\e\es
s  \lb^i &=&  \lx\lb^i
               -f^{ijk} c^j \lb^k
               -\frac{i}{4}\l_{ab}{\bar\s}^{ab}\lb^i
               +i\eta\lb^{i\db}
               %+\frac{3}{2}\kzero\lb^i
               +\frac{1}{2}{\bar \s}^{\m\n}{}\eb {\widetilde F}^i_{\m\n}
               +\frac{i}{2}G^2(\fb_AT^i_{AB}\f_B)\eb\es
s \f_A &=& \lx\f_A
            -T^i_{AB}c^i\f_B
            -\frac{2}{3}i\eta\f_A
            %+\kzero\f_a
            +2\e\p_A\es
s \fb_A &=& \lx\fb_A
             -T^i_{AB}c^i\fb_B
             +\frac{2}{3}i\eta\fb_A
             %+\kzero\fb_a
             -2\eb\pb_A\es
s \p_A &=& \lx\p_A
            -T^i_{AB}c^i\p_B
            +\frac{i}{4}\l_{ab}\p_A\s^{ab}
            +\frac{1}{3}i\eta\p_A
            %+\frac{3}{2}\kzero\p_a
            +i\eb\sbar^\m{\cal D}_\m\f_A
            +2\Lb_{(ABC)}\fb_B\fb_C\e\es
s \pb_A &=& \lx\pb_A
             -T^i_{AB}c^i\pb_B
             -\frac{i}{4}\l_{ab}{\bar\s}^{ab}\pb_A
             -\frac{1}{3}i\eta\pb_A
             %+\frac{3}{2}\kzero\pb_a
             +i\sbar^\m\e{\cal D}_\m\fb_A
             -2 \L_{(ABC)}\f_B\f_C\eb\es
s  c^i &=& \lx c^i
           -\frac{1}{2}f^{ijk}c^jc^k
           -2iE^\m A^i_\m\es
s \cb^i &=& \lx\cb^i
             + b^i
             %+2\kzero\cb^i
             \es
s  b^i &=& \lx b^i
            %+2\kzero b^i
            -2i E^\m\pa_\m\cb^i\ ,
\ea\eqn{brscomplet1}

where $G$ and $\L_{(ABC)}$ are the gauge coupling constant and Yukawa
coupling constants respectively\footnote{ These couplings will
appear in the action displayed only later on (See
\equ{actioninvariante}), although the action cannot be dissociated from
the formulation of the
BRS transformations due to the auxiliary fields having been
eliminated.}.
Note that $\L_{(ABC)}$ is totally
symmetric in its indices.
Moreover, the following notation has been
used:

\eq\ba{lll}
{\widetilde F}^i_{\m\n} &=& F^i_{\m\n}
                 -i\lac \l^i\s_\m\gPb_\n
                       -\l^i\s_\n\gPb_\m
                       -\gP_\n\s_\m\lb^i
                       +\gP_\m\s_\n\lb^i \rac\es
{\cal D}_\m\f_A &=& \nabla_\m\f_A
                    -2i\p_A\gP_\m\es
{\cal D}_\m\fb_A &=& \nabla_\m\fb_A
                    +2i\pb_A\gPb_\m\ .
\ea\eqn{notationsdyn}

Due to the fact that the auxiliary fields of the gauge and matter
multiplets
have been eliminated, the nilpotency of $s$ is true only {\em on-shell},
\ie
\eq
s^2 =\  \mbox{equations of motion for $\l^i$ and $\p_A$}\ .
\eqn{nilpotencebrisee}
(Equations following from the action
\equ{actioninvariante}.)

 For each field transforming nonlinearly in the dynamical fields
we introduce an antifield which  will couple
to its $s$-variation~\cite{brs-gauge,bv}.
This happens for the dynamical fields themselves,
except for $\cb^i$ and $b^i$.

%%%%%%%%%%%%%%%%%%%%%%%%%%%%%%%%%%%%%%%%%%%
\subsection{Flat Limit}\label{lim-plate-poinc}
 Before giving a BRS invariant action and the Slavnov-Taylor
identity, let us have a look at
{\em flat limit} (F.L.) of the curved space-time
model, defined as the limit where we recover the
flat space-time and the rigid superpoincar\'e transformation group.
On the gravitational fields, the flat limit is defined by

\eq\ba{lll}
\vb{\m}{a}&\stackrel{\rm F.L.}{\longrightarrow}&\d_\m^a\es
\gP_\m&\stackrel{\rm F.L.}{\longrightarrow}&0\es
B_\m&\stackrel{\rm F.L.}{\longrightarrow}&0\es
C_{\m\n}&\stackrel{\rm F.L.}{\longrightarrow}&0\ .
\ea\eqn{flchamp}

To be coherent we must specify the flat limit of the ghost in such a way
that the BRS variation of the gravitational fields goes to zero as well,
\ie

\eq
0=s(\fl{\vf}) \stackrel{!}{=} \fl{s\vf}\ ,
\eqn{flvf}

where $\fl{\vf}$ stands for the flat limit of $\vf$.
This is trivial for all the fields except for $\vb{\m}{a}$, which
imposes the
following flat limit for the ghosts:

\eq\ba{lll}
\xi^\m&\stackrel{\rm F.L.}{\longrightarrow}&\fl{\xi^\m}
                                            -\fl{\l^\m {}_a}
x^a\es
\l_{ab}&\stackrel{\rm F.L.}{\longrightarrow}&\fl{\l_{ab}}\es
\e&\stackrel{\rm F.L.}{\longrightarrow}&\fl{\e}\es
\eta&\stackrel{\rm F.L.}{\longrightarrow}&\fl{\eta}\es
w_\m&\stackrel{\rm F.L.}{\longrightarrow}&0\es
w&\stackrel{\rm F.L.}{\longrightarrow}&0\ ,
\ea\eqn{flghosts}

where $\fl{\xi^\m}, \fl{\l_{ab}}, \fl{\e}, \fl{\eta}$ are the
constant ghosts used in the rigid model.

\underline{Remark}: We see that when going from flat space-time
to curved space-time, and thus from global Poincar\'e
to local Poincar\'e transformations,
the orbital part of the Lorentz transformation goes to the
diffeomorphisms
and only the spin part stays in the local Lorentz transformation.

\subsection{Invariant Action}

An action which is invariant under $s$ is given by

\eq
\S_{\rm inv} = \S_{\rm SYM} +\S_{\rm gf}\ .
\eqn{actioninvariante}

In order to write down $\S_{\rm SYM}$,
let us introduce for convenience the
following filtration operator:

\eq
N=\intx\lac \gP_\m \dfud{}{\gP_\m}
            -\gPb_\m\dfud{}{\gPb_\m}
            +C_{\m\n}\dfud{}{C_{\m\n}}\rac\ ,
\eqn{filtration}
which allows us to write
\eq
\S_{\rm SYM} =\sum_{i=0}^4\S_{\rm SYM}^{(i)}\ ,
\hspace{2cm} \mbox{with}\hspace{1cm}
N\S_{\rm SYM}^{(i)}=i\S_{\rm SYM}^{(i)}\ .
\eqn{actionfiltree}

Explicitly, we then have

\eq\ba{rl}
\S_{\rm SYM}^{(0)} = \intx \sqrt{g}\LP &\!\!\!
    \displaystyle{
    -\frac{1}{4G^2}F^{i\m\n}F^i_{\m\n}
    -\frac{i}{G^2}\l^{i}\s^\m\nabla_\m\lb^i
    -\frac{G^2}{8} |\fb_AT^i_{AB}\f_B|^2
    +\frac{1}{2}\nabla_\m\f_A\nabla^\m\fb_A}
\\
&\!\!\!
    \displaystyle{
  -i\p_A\s^\m\nabla_\m\pb_A
  -2\L_{(ABC)}\f_B\f_C\Lb_{(ADE)}\fb_D\fb_E
  -iT^i_{AB}\f_B\pb_A\lb^i}
\\
&\!\!\!
    \displaystyle{
  -iT^i_{AB}\fb_A\p_B\l^i
  +2\L_{(ABC)}\p_A\p_B\f_C
  +2\Lb_{(ABC)}\pb_A\pb_B\fb_C
  -\frac{1}{12} R\f_A\fb_A}
\RP
\ea\eqn{SYM0}

\eq\ba{rl}
\S_{\rm SYM}^{(1)} = \intx \sqrt{g}\LP &\!\!\!
    \displaystyle{
    -\frac{1}{2G^2}\lp \l^i\s^\m\sbar^{\n\r}\gPb_\m +
                         \gP_\m\s^{\n\r}\s^\m\lb^i \rp F^i_{\n\r}
    }
\\
&\!\!\!
    \displaystyle{
    -\frac{i}{2}\lp\fb_A T^i_{AB}\f_B\rp\lp\gP_\m\s^\m\lb^i
    +\l^i\s^\m\gPb_\m\rp    }
\\
&\!\!\!
    \displaystyle{
    -i\lp\p_A\s^\m\sbar^\n\gP_\m\rp\nabla_\n\fb_A
    +i\lp\gPb_\m\sbar^\n\s^\m\pb_A\rp\nabla_\n\f_A }
\\
&\!\!\!
    \displaystyle{
    -\frac{2}{3}\lp\pb_A\sbar^{\m\n}\nabla_\m\gPb_\n\rp\f_A
    -\frac{2}{3}\lp\nabla_\m\gP_\n\s^{\m\n}\p_A\rp\fb_A
}
\\
&\!\!\!
    \displaystyle{
    +2\L_{(ABC)}\lp\p_A\s^\m\gPb_\m\rp\f_B\f_C
    +2\Lb_{(ABC)}\lp\gP_\m\s^\m\pb_A\rp\fb_B\fb_C
    }
\RP
\ea\eqn{SYM1}

\eq\ba{rl}
\S_{\rm SYM}^{(2)} = \intx \sqrt{g}\LP &\!\!\!
    \displaystyle{\frac{1}{G^2}\biggl\{
    -\frac{3}{8}V^\m\lp\l^i\s_\m\lb^i\rp
    +\lp\l^i\s^\l\gPb_\n\rp\lp\l^i\s^\n\gPb_\l\rp
    }
\\
&\!\!\!
    \displaystyle{
    +\lp\gP_\l\s^\n\lb^i\rp\lp\gP_\n\s^\l\lb^i\rp
    +\frac{1}{2}\lp\l^i\s^\n\gPb_\n\rp\lp\l^i\s^\l\gPb_\l\rp
    }
\\
&\!\!\!
    \displaystyle{
    +\frac{1}{2}\lp\gP_\l\s^\l\lb^i\rp\lp\gP_\n\s^\n\lb^i\rp
    +\frac{1}{2}\lp\l^i\s^\l\gPb_\n\rp\lp\gP_\l\s^\n\lb^i\rp
    }
\\
&\!\!\!
    \displaystyle{
    +\frac{1}{2}\lp\l^i\s^\n\gPb_\n\rp\lp\gP_\l\s^\l\lb^i\rp
    -\lp\gP_\l\s^\n\lb^i\rp\lp\l^i\s_\n\gPb^\l\rp
    }\biggr\}
\\
&\!\!\!
    \displaystyle{
    +\frac{1}{8} X^\m \lp\p_A\s_\m\pb_A\rp
    +2\lp\p_A\gP_\m\rp\lp\pb_A\gPb^\m\rp
    }
\\
&\!\!\!
    \displaystyle{
    +\frac{2i}{3}\Lb_{(ABC)}\fb_A\fb_B\fb_C\lp\gP_\m\s^{\m\n}\gP_\n\rp
    +\frac{2i}{3}\L_{(ABC)}\f_A\f_B\f_C\lp\gPb_\m\sbar^{\m\n}\gPb_\n\rp
    }
\\
&\!\!\!
    \displaystyle{
    +\frac{1}{6}\f_A\fb_A\nabla_\m F_\n{}^{\m\n}
    +\frac{i}{8}X^\m\lp\f_A\nabla_\m\fb_A-\nabla_\m\f_A\fb_A\rp
    }
\\
&\!\!\!
    \displaystyle{
    +\frac{1}{3}\f_A\fb_A\e^{\m\n\r\l}\lp\gP_\n\s_\l\nabla_\m\gPb_\r
                                      -\nabla_\m\gP_\n\s_\l\gPb_\r\rp
    }
\RP
\ea\eqn{SYM2}

\eq\ba{rl}
\S_{\rm SYM}^{(3)} = \intx \sqrt{g}\LP &\!\!\!
    \displaystyle{
    \f_A\lp\pb_A\gPb_\m\rp\lp -\frac{1}{4}V^\m+\frac{1}{3}Y^\m
        -\frac{i}{3}F_\n {}^{\m\n}\rp
    }\\ &\!\!\!
    \displaystyle{
    +\fb_A\lp\p_A\gP_\m\rp\lp -\frac{1}{4}V^\m+\frac{1}{3}Y^\m
        +\frac{i}{3}F_\n {}^{\m\n}\rp
    }
\RP
\ea\eqn{SYM3}

\eq\ba{rl}
\S_{\rm SYM}^{(4)} = \intx \sqrt{g}\LP &\!\!\!
    \displaystyle{
    \frac{1}{32}\f_A\fb_A V_\m V^\m
    +\frac{1}{12}\f_A\fb_A\lac F_\r {}^{\m\r} F_{\n\m}{}^\n
                              +F^{\n\m\r} F_{\m\r\n}\rac
    }
\RP\ ,
\ea\eqn{SYM4}

and, for the gauge fixing part -- of the Landau type:

\eq\ba{rl}
\S_{\rm gf} = s\intx \sqrt{g}&\lac
 -\pa_\m\cb^i g^{\m\n} A^i_\n\rac\es
            = \phantom{s} \intx \sqrt{g}&\lac
 -2\lp\e\s^\l\gPb_\l +\gP_\l\s^\l\eb\rp\pa_\m\cb^i g^{\m\n} A^i_\n
 -\pa_\m b^i g^{\m\n} A^i_\n\right.\es &\left.
 -2\pa_\m \cb^i\lp\e\s^\m\gPb^\n
                  +\e\s^\n\gPb^\m
                  +\gP^\m\s^\n\eb
                  +\gP^\n\s^\m\eb\rp A^i_\n\right.\es &\left.
 +\pa_\m \cb^i g^{\m\n} \lp\nabla_\n c^i +\e\s_\n\lb^i +
\l^i\s_\n\eb\rp\rac\ .
\ea\eqn{gfaction}

\underline{Remark}:  This action is the most general power-counting
renormalizable\footnote{The most general invariant action, with
the power-counting condition relaxed, may be found in~\cite{brandt94}.}
one invariant under $s$, however up to purely gravitational terms,
such as the purely supergravity action, which we have omitted.

\subsection{Classical Identities Defining the Model}
\label{cidtm}

As usually in the framework  of algebraic
renormalization~\cite{pi-so}, we want to define the theory through
a complete set of functional identities. These identities express the
various symmetries of the model as well as specific properties
linked to the
choice made for the gauge fixing. The action \equ{actioninvariante}
ought to be the general solution of these functional identities, up to
field and parameter redefinitions.

\subsubsection{ The Slavnov-Taylor (ST) Identity}

To write down the ST identity --  the Ward identity of
BRS invariance --
we add to the invariant action the following
external action $\S_{\rm ext}$ which contains, besides the usual
antifields
coupled to the nonlinear $s$-variations of the dynamical fields,
also non-standard terms, quadratic in the antifields, in order
to correct
for the fact that the $s$-operator  was on-shell rather than off-shell
nilpotent:

\eq
\S_{\rm ext} = \intx\lp\dsum{\vf\in I}\vf^* s\vf
-\frac{G^2}{2\sqrt{g}}\lp\e\l^{*i}-\eb\lb^{*i}\rp^2
+\frac{2}{\sqrt{g}}\lp\e\p^*_A\rp\lp\eb\pb^*_A\rp
                 \rp \ ,
\eqn{actionexterieure}

where $I$ is defined below, in (\ref{defI}).
Then, for the total action $\S$,

\eq
\S= \S_{\rm inv}+\S_{\rm ext}\ ,
\eqn{action totale}

we have the ST identity

\eq
S(\S)=0\ ,
\eqn{STidentity}

with

\eq
S(\g )\equiv \intx \lac \dsum{\vf\in I}
      \dfud{\g}{\vf^*}\dfud{\g}{\vf}
      +\dsum{\vf '\in I'} s\vf ' \dfud{\g}{\vf '} \rac\ ,
\eqn{operateurdeST}

Here and throughout the paper, we define the three sets of fields

\eq\ba{l}
I=\lac A_\m^i ,\l^i ,\p_A ,\f_A ,c^i \rac\ , \quad
I^*=\lac A^{*i}_\m ,\l^{*i} ,\p^*_A ,\f^*_A ,c^{*i} \rac\ ,\es
I' =\lac \cb^i ,b^i  \
\mbox{and the supergravity fields and ghosts}\,\rac\ .
%\vb{\m}{a},
%\gP_\m , C_{\m\n},B_\m ,\xi^\m ,\l_{ab} ,\e ,\eta ,w_\m ,w\rac\ ,
\ea\eqn{defI}

containing the dynamical fields which transform
nonlinearly, the corresponding antifields and
fields which transform linearly, including the gravitational fields
and  ghosts, respectively.
For further use, we introduce the Batalin-Vilkovisky~\cite{bv}
notation for the ST operator:

\eq
S(\g)=\frac{1}{2}(\g ,\g)+b\g\ ,
\eqn{BVnotation}

with

\eq\ba{ll}
\displaystyle{
(\g ,\g ')\equiv \intx \lac \dsum{\vf\in I}\lp
      \dfud{\g}{\vf^*}\dfud{\g '}{\vf}
      +\dfud{\g '}{\vf^*}\dfud{\g}{\vf} \rp\rac\ ,}
\hspace{5mm}&\forall\g,\g'\hspace{4mm}
\mbox{with}\hspace{4mm}GP(\g)=GP(\g')=0\ ,
\es
\displaystyle{
b\equiv \intx \lac \dsum{\vf '\in I'} s\vf ' \dfud{}{\vf '} \rac \ ,}
&\mbox{with}\hspace{4mm}b^2=0\ .
\ea\eqn{BVnotation2}

 From (\ref{STidentity}), we deduce that the linearized ST
operator

\eq
\BB_\S = (\S,\cdot )+b\ ,
\eqn{STlinearise}
is nilpotent:

\eq
\BB_\S^2 = 0\ .
\eqn{nilpotencedeSTlinearise}

\subsubsection{Gauge Condition and Ghost  Identities}

Besides the ST identity, we need to impose other identities:

{\bf The gauge condition}

\eq
\dfud{\S}{b^i(x)}=\O_b^i(x) \ ,
\eqn{gaugecondition}

where $\O_b^i = \pa_\m\lp \sqrt{g} g^{\m\n} A^i_\n\rp$ is a classical
breaking, \ie is linear in the dynamical fields.  It defines the Landau
gauge in curved space-time.

{\bf The ghost identities}

\eq
F^i\S=\D_{\rm G}^i\ ,\hspace{3cm}
\mbox{with}\hspace{1cm} F^i = \intx \lac \dfud{}{c^i}
                 -f^{ijk}\cb^j\dfud{}{b^k}\rac\ .
\eqn{ghostequation}

This identity is peculiar to the Landau gauge~\cite{bps} and
$\D_{\rm G}^i$ is a classical breaking
given in \ref{appendixd}. Moreover:

\eq\ba{lll}
\dfud{\S}{\xi^\m(x)}=\O_{{\rm D}\m}(x)\ ,\hspace{1cm}&
\dfud{\S}{\l_{ab}(x)}=\O^{ab}_{{\rm L}}(x)\ ,\hspace{1cm}&
\dfud{\S}{\eta(x)}=\O_{\rm R}(x)\ ,\hspace{1cm}\es
\dfud{\S}{w_\m(x)}=0\ ,&
\dfud{\S}{w(x)}=0\ ,
\ea\eqn{ghostequations}

where $\O_{{\rm D}\m}, \O^{ab}_{{\rm L}}$ and $\O_{\rm R}$ are classical
breakings given in \ref{appendixd}.  The equations \equ{ghostequations}
follow from the  diffeomorphisms, Lorentz transformations and
$R$-transformations being linear and from the dynamical fields being
invariant under the $C$-transformations.

By ``commuting'' the
identities (\ref{gaugecondition}-\ref{ghostequations}) with the ST
identity (\ref{STidentity}), we recover the antighost equation, the
$\GG$-rigid invariance, and the  Ward identities for the
invariance under diffeomorphisms,
Lorentz transformations, $R$- and $C$-transformations.
They are explicitly given in \ref{appendixd}.
This shows that all these symmetries are encoded in the ST identity,
 the gauge condition and
the ghost identities, which constitute the minimal set of identities
defining the model.

\subsection{Supercurrent and Associated Ward Identities}

Let us finish this section by explicitly giving the supercurrent components
and conservation laws of the model.

\subsubsection*{Supercurrent Components:}
We identify the components of the supercurrent multiplet as the
insertions

\eq
R^\mu = \dfud{\S}{B_\m}\ ,\quad Q^\m{}_\a = \dfud{\S}{\Psi^\a_\m}\ ,
\quad T^\m{}_\n = e_\n{}^a\dfud{\S}{e_\m{}^a}\ .
\eqn{currents}

That they belong to a supermultiplet  is encoded in the $\e$ part
-- the proper supersymmetry part -- of the
transformation laws of the supergravity fields.

\subsubsection*{Conservation Laws:}
We readily see that the Ward identities for local
$R$-invariance (last of \equ{symmlin}, with \equ{omegar}) and for the
diffeomorphisms (third of \equ{symmlin}, with \equ{omegadiff1})
correspond to the conservation of the
current $R^\m$ and of the energy-momentum tensor $T^\m{}_\n$:

\eq\ba{ll}
\pa_\m R^\m = &\sumall iR_\vf\vf\dfud{\S}{\vf}
+\pa_\m\lp\xi^\m\dfud{\S}{\eta}\rp
\ ,\es
\pa_\m T^\m{}_\n =  &\sumall \pa_\n\vf\dfud{\S}{\vf}
+ \pa_\m \LP\d_\n^\m\displaystyle{\sum_{\vf^*\in I^*}\vf^*\dfud{}{\vf^*}}
-A^i_\n\dfud{}{A^i_\m} + A^{*i\m}\dfud{}{A^{*i\n}}
%\es&
-\gP_\n\dfud{}{\gP_\m}
+\gPb_\n\dfud{}{\gPb_\m}
\es&
-C_{\n\r}\dfud{}{C_{\m\r}}
-B_\n\dfud{}{B_\m}
-w_\n\dfud{}{w_\m}
\RP \S
+\pa_\m\lp\xi^\m\dfud{\S}{\xi^\n}\rp
\ .
\ea\eqn{R-T-conserv}

The set $I^*$ in the summation has been defined
in \equ{defI}.

Moreover, we can use the local Lorentz
invariance (fourth of \equ{symmlin}, with \equ{omegalorentz2})
to show that this conserved energy-momentum tensor
is symmetric. Indeed, its antisymmetric part vanishes on-shell:

\eq\ba{rl}
T^{[ab]}=&
\frac{i}{4}\l^i\s^{ab}\dfud{\S}{\l^i}
-\frac{i}{4}\lb^i\s^{ab}\dfud{\S}{\lb^i}
+\frac{i}{4}\p_A\s^{ab}\dfud{\S}{\p_A}
-\frac{i}{4}\pb_A\s^{ab}\dfud{\S}{\pb_A}
+\frac{i}{4}\gP_\m\s^{ab}\dfud{\S}{\gP_\m}
\es &
-\frac{i}{4}\gPb_\m\s^{ab}\dfud{\S}{\gPb_\m}
+\frac{i}{4}\e\s^{ab}\dfud{\S}{\e}
-\frac{i}{4}\eb\s^{ab}\dfud{\S}{\eb}
+          \lp\d^a_c\l^b{}_d
              -\d^b_c\l^a{}_d\rp\dfud{\S}{\l_{cd}}
\es &
+\pa_\m\lp\frac{1}{2}\xi^\m\dfud{\S}{\l_{ab}}\rp\ .
\ea\eqn{improvedtmn}

As we will see later, $T^\m{}_\n$ is indeed the
improved energy-momentum tensor, as its trace $T^\m{}_\m$ vanishes
as well on-shell \equ{supertraceids}.

For the supersymmetry current  $Q^\m{}_\a$,
and in the F.L.,
we use the local
supersymmetry Ward identity \equ{susywi}, which yields

\eq
\left.\lp\pa_\m Q^\m{}_\a\rp\right|_{{\rm ghosts}=0}
= i\left.\lp \S,\dfud{\S}{\e^\a} \rp\right|_{{\rm ghosts}=0}
    + \mbox{ terms vanishing in the flat limit} \ .
\eqn{Q-conserv}

\section{Weyl, Supertrace and Shift Identities}

In this section, we will exhibit other symmetries of the action $\S$
which are of particular interest for the purpose of our work.
They are the Weyl symmetry~\cite{weylsym,frad-tsey,kraus-sib},
the so-called supertrace
identities~\cite{fer-zum,cps-supertr,ps-book} which can now be
written as a Ward identity thanks
to
the introduction of the external gravitational fields, and finally a new
identity, called the shift identity, which links the $C_{\m\n}$ and
$B_\m$ dependence of the action. This last symmetry, as we will see,
must be
introduced in order to have a closed algebra.

\subsection{Weyl Identity}

The Weyl transformations of infinitesimal {\em local} parameter $\k(x)$
are~\cite{weylsym,frad-tsey,kraus-sib}
\eq
\d_\k \vf(x) = \wtd_\vf \k(x)\vf(x)\ ,
\eqn{Weyltransf}

for all fields $\vf$ of the model. The Weyl weights $\wtd_\vf$ are
given by

\eq\ba{ll}
\wtd_\vf = d_\vf
-\mbox{number of covariant indices $\m$ of $\vf_\m$}\hspace{1cm}&
\es
\phantom{\wtd_\vf = d_\vf}
+\mbox{number of contravariant indices $\m$ of $\vf^\m$,}\hspace{1cm}&
\mbox{for}\;\ \vf\in I\cup I'\ ,\es
\wtd_{\vf^*} = -\wtd_\vf\ ,&
\mbox{for}\;\ \vf\in I\ ,
\ea\eqn{weylweights}

the sets $I$ and $I'$ being defined by \equ{defI}
and $d_\vf$ being the canonical dimension of $\vf$.
For further use, we write the associated Ward operator

\eq
S_{\rm Weyl} = \intx\lac
\sumall\wtd_\vf \k \vf\dfud{}{\vf}\rac\ .
\eqn{weyloperator}

The Weyl transformations are in fact the generalization to the curved
space-time formalism of the dilatation
transformations~\cite{weylsym,frad-tsey,kraus-sib}.

By a direct calculation, we find

\eq
S_{\rm Weyl}\S_{\rm SYM} =
-\frac{3}{4}\intx\lac
\e_\m {}^{\n\r\l}\pa_\n\k C_{\r\l}\dfud{\S_{\rm SYM}}{B_\m}\rac\ .
\eqn{weylsymmetry}

\underline{Remark}: We see from (\ref{weylsymmetry}) that although
$\S_{\rm SYM}$ is not invariant under {\em local} Weyl transformations,
it is invariant under {\em global} Weyl transformations, as the r.h.s.
of (\ref{weylsymmetry}) vanishes for a constant infinitesimal
parameter $\k$.
This is directly linked to the dilatation invariance of the flat limit
model.

Thanks to the operational form of the r.h.s of
(\ref{weylsymmetry}) we can
define the following modified operator

\eq
S_\k = S_{\rm Weyl} +\frac{3}{4}\intx\lac
\e_\m {}^{\n\r\l}\pa_\n\k C_{\r\l}\dfud{}{B_\m}\rac\ ,
\eqn{skappaoperator}

to obtain

\eq
S_\k \S_{\rm SYM} = 0\ .
\eqn{skappainvariance}

Next, applying $S_\k$ on $\S_{\rm ext}$ and $\S_{\rm gf}$ leads to:

\eq\ba{ll}
S_\k\S_{\rm ext}=\D_{\k ,{\rm ext}}\equiv\intx\LP &\!\!\!
-i\pa_\m\k\lp\p^*_A\s^\m\eb\f_A-\pb^*_A\sbar^\m\e\fb_A\rp
\es&\!\!\!
+\frac{1}{2}\e_\m{}^{\n\r\l}\pa_\n\k C_{\r\l}
\lp\p^*_A\s^\m\eb\f_A+\pb^*_A\sbar^\m\e\fb_A\rp
\es&\!\!\!
+\pa_\m\k\ \xi^\m \lp
\displaystyle{\sum_{\vf\in I}}(-1)^{{\rm gh}(\vf^*)}\ \wtd_\vf\vf^*\vf
\rp
\RP\ ,
\ea\eqn{brisurekappaext}
\eq
S_\k\S_{\rm gf}=\D_{\k ,{\rm gf}}\equiv\intx\LP
-2\pa_\m\k\lac
\BB_\S\lp
\sqrt{g}\cb^i g^{\m\n}A^i_\n\rp
 -\lx\lp\sqrt{g}\cb^i g^{\m\n}A^i_\n\rp
\rac\RP\ ,
\eqn{brisurekappagf}

and thus

\eq
S_\k\S = \D_{\k ,{\rm ext}} + \D_{\k ,{\rm gf}}\ .
\eqn{brisurekappa}

Again, we note that $\S$, although not invariant under the local
modified
Weyl transformations, is invariant under the global Weyl
transformations.

It should be stressed that, as it stands, (\ref{brisurekappa}) cannot be
the
starting point of the study of the quantum case since $\D_{\k ,{\rm
gf}}$
is nonlinear in the dynamical fields and thus needs to be correctly
defined
at the quantum level ($\D_{\k ,{\rm ext}}$, being linear in the
dynamical
fields, is well defined in the quantum model). Although we are only
interested
in the classical model in this paper, it is only the first step towards
a
quantum treatment  and thus a suitable form of
(\ref{brisurekappa})
is needed. We will return to this problem later (see  Section
(\ref{gaugefixingproblem})),
where we will introduce sources  in order to define the operator
$\lp\sqrt{g}\cb^i g^{\m\n}A^i_\n\rp$ and its
$\BB_\S$-variation at the quantum level.

 \underline{Remark}: As we see from the Ward operator
\equ{skappaoperator}, there was no need to introduce a proper gauge
field associated to the local Weyl invariance. In a certain sense, it
is $B_\m$ -- the gauge field of local $R$-invariance -- which plays
this role.

%%%%%%%%%%%%%%%%%%%%%%%%%%%%%%%%%%%%%%%%%%%%%%%%%%
\subsection{Supertrace Identities}
 It is known~\cite{fer-zum,cps-supertr,pigsor,pigsib,ps-book}
 that
the trace of the energy-momentum tensor, $T^\m{}_\m$, the trace of the
supersymmetry current,  $\s^\m_{\a\da}{\bar Q}^\da_\m$, and the
divergence of the
$R$-current, $\pa_\m
R^\m$ -- all vanishing on shell in the massless classical
theory considered here -- belong to a supermultiplet, and that the
corresponding trace and conservation identities form a supermultiplet of
equations: the ``supertrace identities''.

 Our starting point will be the trace identity for the supersymmetry
current, the other identities being deduced from it by
covariance under supersymmetry.
For the theory defined by the SYM  action
\equ{actionfiltree}, it reads:

\eq
\s^\m\dfud{\S_{\rm SYM}}{\gPb_\m}=
2\f_A\dfud{\S_{\rm SYM}}{\p_A}
+3 \s^\n\sbar_\m\gP_\n\dfud{\S_{\rm SYM}}{B_\m}\ .
\eqn{supertrace1}

 Introducing the infinitesimal local parameter $\chi_\a(x)$ -- a Weyl
spinor -- we may
rewrite this equation as an integral Ward identity:
\eq
\widetilde{\widetilde{S}}_\chi \S_{\rm SYM} = 0\ ,
\eqn{tilde-wi}

 where we have defined the  Ward operator
$\widetilde{\widetilde{S}}_\chi$ by

\eq
\widetilde{\widetilde{S}}_\chi = \intx\lac
-\chi\s_\m\dfud{}{\gPb_\m}
+2\f_A\chi\dfud{}{\p_A}
+3 \lp\chi\s^\n\sbar_\m\gP_\n\rp\dfud{}{B_\m}\rac\ .
\eqn{supertrace2}

Applying $\widetilde{\widetilde{S}}_\chi$ to $\S_{\rm ext}$
\equ{actionexterieure}
leads to:

\eq\ba{rl}
\widetilde{\widetilde{S}}_\chi\S_{\rm ext} = \intx\LP&\!\!\!
-\lp\l^{*i}\s^{\m\n}\e\rp\lp\chi\s_{\m\n}\l^i\rp
+4\lp\chi\e\rp\lp\pb_A\pb^*_A+\f_A\f^*_A\rp\es &\!\!\!
+4\fb_A\lp\e\gP_\m\rp\lp\chi\s^\m\pb^*_A\rp
-4\f_A\lp\chi\p^*_A\rp\lp\gP_\m\s^\m\eb\rp\es &\!\!\!
-2\f_A\chi\pa_\m\lp\xi^\m\p^*_A\rp
-2\lp\chi\p^*_A\rp T^i_{AB}\f_Bc^i\es &\!\!\!
+\frac{i}{2}\f_A\lp\chi\s^{ab}\p^*_A\rp\l_{ab}
+\frac{2i}{3}\f_A\lp\chi\p^*_A\rp\eta
\RP\es
= \intx\LP&\!\!\!
2\lp\chi\p^*_A\rp\dfud{\S}{\f^*_A}
+i\lp\chi\s_{ab}\e\rp\dfud{\S}{\l_{ab}}
-3i\lp\chi\e\rp\dfud{\S}{\eta}\es &\!\!\!

%-2\lp\chi\e\rp\lp
%\frac{3}{2}\l^{*i}\l^i
%+\frac{3}{2}\lb^{*i}\lb^i
%-\f^*_A\f_A
%-\fb^*_A\fb_A
%+\frac{3}{2}\p^*_A\p_A
%+\frac{3}{2}\pb^*_A\pb_A\rp
-2\lp\chi\e\rp\lp
\displaystyle{\sum_{\vf\in I}}(-1)^{{\rm gh}(\vf^*)}\ \wtd_\vf\vf^*\vf
\rp \vspace{4mm}
\es &\!\!\!
+2i\f_A\lp\chi\p^*_A\rp\eta
+\frac{i}{2}\f_A\lp\chi\s^{ab}\p^*_A\rp\l_{ab}
-2\chi\pa_\m\lp\xi^\m\p^*_A\f_A\rp\es &\!\!\!
+4\fb_A\lp\e\gP_\m\rp\lp\chi\s^\m\pb^*_A\rp
-4\f_A\lp\chi\p^*_A\rp\lp\gP_\m\s^\m\eb\rp
\RP\ ,
\ea\eqn{supertrace3}

Thus redefining $\widetilde{\widetilde{S}}_\chi$ by
\eq\ba{ll}
\widetilde{\widetilde{S}}_\chi\rightarrow
{\widetilde{S}}_\chi= \intx&\!\!\!\lac
-\chi\s_\m\dfud{}{\gPb_\m}
+2\f_A\chi\dfud{}{\p_A}
+3\lp\chi\s^\n\sbar_\m\gP_\n\rp\dfud{}{B_\m}
\right.\es &\!\!\!\left.
-2\lp\chi\p^*_A\rp\dfud{}{\f^*_A}
-i\lp\chi\s_{ab}\e\rp\dfud{}{\l_{ab}}
+3i\lp\chi\e\rp\dfud{}{\eta}
\rac\ ,
\ea\eqn{redefschi}

and defining  the classical breaking $\D_{\chi ,{\rm ext}}$ by

\eq\ba{rl}
\D_{\chi ,{\rm ext}} = \intx\LP&\!\!\!
-2\lp\chi\e\rp\lp
\displaystyle{\sum_{\vf\in I}}(-1)^{{\rm gh}(\vf^*)}\ \wtd_\vf\vf^*\vf
\rp\vspace{4mm}\es &\!\!\!
+2i\f_A\lp\chi\p^*_A\rp\eta
+\frac{i}{2}\f_A\lp\chi\s^{ab}\p^*_A\rp\l_{ab}
-2\chi\pa_\m\lp\xi^\m\p^*_A\f_A\rp\es &\!\!\!
+4\fb_A\lp\e\gP_\m\rp\lp\chi\s^\m\pb^*_A\rp
-4\f_A\lp\chi\p^*_A\rp\lp\gP_\m\s^\m\eb\rp
\RP\ ,
\ea\eqn{supertrace3bis}

allows to write

\eq
{\widetilde{S}}_\chi\S_{\rm ext}=\D_{\chi ,{\rm ext}}\ .
\eqn{supertrace3ter}

Then, applying the modified operator on $\S_{\rm gf}$ leads to:

\eq
{\widetilde{S}}_\chi\S_{\rm gf} %\equiv\D_{\chi ,{\rm gf}} =
=\D_{\chi ,{\rm gf}}+
\intx\LP
%4\pa_\m\lp\chi\e\rp\lp\sqrt{g}\cb^i g^{\m\n}A^i_\n\rp
%+
4\lp\chi\e\rp\cb^i\dfud{\S_{\rm gf}}{b^i}
\RP\ ,
\hspace{15mm}\mbox{with}\hspace{5mm}
\D_{\chi ,{\rm gf}}= \intx\LP
4\pa_\m\lp\chi\e\rp\lp\sqrt{g}\cb^i g^{\m\n}A^i_\n\rp
\RP\ .
\eqn{supertrace4}

We see again the appearance of the operator
$\lp\sqrt{g}\cb^i g^{\m\n}A^i_\n\rp$ --  quadratic
in the dynamical fields -- in the breaking of the
supertrace identity, which deserves a better insight, postponed
to section (\ref{gaugefixingproblem}). We can redefine
once more the
operator ${\widetilde{S}}_\chi$,
absorbing the second term of the r.h.s. of (\ref{supertrace4}) in it:
\eq\ba{ll}
\mbox{}{\widetilde{S}}_\chi\rightarrow
S_\chi= \intx&\!\!\!\lac
-\chi\s_\m\dfud{}{\gPb_\m}
+3 \lp\chi\s^\n\sbar_\m\gP_\n\rp\dfud{}{B_\m}
+2\f_A\chi\dfud{}{\p_A}
-2\lp\chi\p^*_A\rp\dfud{}{\f^*_A}
\right.\es &\!\!\!\left.
-i\lp\chi\s_{ab}\e\rp\dfud{}{\l_{ab}}
+3i\lp\chi\e\rp\dfud{}{\eta}
- 4\lp\chi\e\rp\cb^i\dfud{}{b^i}
\rac\ .
\ea\eqn{reredefschi}

Thus

\eq
S_\chi\S = \D_{\chi ,{\rm ext}}+\D_{\chi ,{\rm gf}}\ .
\eqn{wardchi}

For further use, let us denote by ${\bar S}_\chib $ the complex
conjugate
of $S_\chi$ and by
${\bar\D}_{\chib ,{\rm ext}}$ and ${\bar\D}_{\chib ,{\rm gf}}$
the complex conjugates of $\D_{\chi ,{\rm ext}}$ and $\D_{\chi ,{\rm
gf}}$,
respectively.

\underline{Remark}: The supertrace operator $S_\chi$ acts on the
supergravity and matter fields, but not on the gauge fields.

%%%%%%%%%%%%%%%%%%%%%%%%%%%%%%%%%%%%%%%%%%%%%%%%%%%%
\subsection{Shift  Identity}

Having written the new Ward identities (\ref{brisurekappa}) and
(\ref{wardchi})
(and their complex conjugates),
we have now to study the algebra they form with the identities
defining the theory,
and particularly with the ST identity (\ref{STidentity}).
Thus, for any $\g$ with zero GP, we have

\eq\ba{ll}
S_\chi S(\g) - \BB_\g\lp S_\chi\g-\D_{\chi ,{\rm ext}}-\D_{\chi ,{\rm
gf}}\rp
&= \left[ S_\chi,b\right]\g + \lp \g,\D_{\chi ,{\rm ext}}\rp
+b\D_{\chi ,{\rm ext}} + \BB_\g \D_{\chi ,{\rm gf}} \es
&= {\dot S}_\chi\g+b\D_{\chi ,{\rm ext}} + \BB_\g \D_{\chi ,{\rm gf}}\ ,
\ea\eqn{commslavchi}
where we have used the definitions
(\ref{BVnotation2},\ref{STlinearise}), and with
\eq\ba{ll}
{\dot S}_\chi &
= S_\k \left[-2\lp\chi\e\rp\right]
+ S_K \left[-4\lp\chi\e\rp C_{\m\n}-4i\lp\e\s_{\m\n}\chi\rp\right]\es &
+ S_\chi \left[\lx\chi+\frac{i}{4}\l_{ab}\chi\s^{ab}+i\eta\chi
  -2\lp\gP_\m\s^\m\eb\rp\chi\right]
+ {\bar S}_\chib \left[-2\lp\e\gP_\m\rp\chi\s^\m\right]
\es &
+\intx\lac
-12i \lp\chi\s^\m\eb\rp\lp\e\gP_\m\rp\dfud{}{\eta}
-4\lp\chi\e\rp w_\m\dfud{}{w_\m}
-4\lp\chi\e\rp w\dfud{}{w}\rac\ ,
\ea\eqn{dotchi}
where we have introduced the notation
$S_\a[\ldots]$ for the Ward operator
obtained by replacing the infinitesimal parameter
$\ \a=\chi,\chib,\k\ $ by the argument
specified in the brackets, and have
defined, for any antisymmetric tensor $K_{\m\n}$:
\eq
S_K = \intx\lac \frac{1}{2}K_{\m\n}\dfud{}{C_{\m\n}}
-\frac{3}{8}\e_\m{}^{\n\r\l}\pa_\n K_{\r\l}\dfud{}{B_\m}\rac\ .
\eqn{defsk}

Setting $\g = \S$ in(\ref{commslavchi}), we get

\eq
{\dot S}_\chi \S +b\D_{\chi ,{\rm ext}} + \BB_\S\D_{\chi ,{\rm gf}}= 0\
.
\eqn{dotssigma}

Next, using (\ref{dotchi}), (\ref{brisurekappa}),
(\ref{wardchi}) and (\ref{ghostequations}), we obtain the following
 (linearly broken) {\it shift identity}\footnote{This
identity can be
 directly checked on the action $\S$ given by \equ{action totale}.}
\eq
S_K \S = \D_{K, {\rm ext}}\ ,
\eqn{wardk}

where the classical breaking $\D_{K, {\rm ext}}$ is given by

\eq
\D_{K, {\rm ext}} =
\intx \lp -\frac{1}{4}\e_\m{}^{\n\r\l}\pa_\n K_{\r\l}
\lac \p^*_A\s^\m\eb \f_A
     + \pb^*_A\sbar^\m\e \fb_A\rac\rp\ .
\eqn{classbrk}

As the Ward operator $S_K$ appears in the r.h.s of (\ref{commslavchi}),
its presence is necessary for closing the algebra.

\underline{Remark}: The denomination of (\ref{wardk})
as the shift identity can be justified by defining

\eq
{\hat B}_\m = B_\m +\frac{3}{8} \e_\m{}^{\n\r\l}\pa_\n C_{\r\l}\ ,
\eqn{defhatb}

and thus getting, from (\ref{wardk})

\eq
\S(B_\m, C_{\m\n}, \vf)=\S({\hat B}_\m, 0, \vf)
+ \left.\D_{K,{\rm ext}}\right|_{K_{\m\n}=C_{\m\n}}\ ,
\eqn{shift}

which shows that the action, except the ``classical'' term
$\D_{K,{\rm ext}}|_{K=C}$,
depends on $C_{\m\n}$ only as a {\em shift}
of $B_\m$.

\subsection{Algebra}

Let us finally show that the algebra formed by
$S_\k , S_\chi , {\bar S}_\chib , S_K,$ the gauge condition
(\ref{gaugecondition}),
the ghost identities (\ref{ghostequation}-\ref{ghostequations})
and the ST identity (\ref{STidentity})
is closed, by exhibiting the nontrivial ''commutation" rules. Thus,
for any $\g$ with zero GP, we have

\eq\ba{ll}
S_K S(\g) - \BB_\g\lp S_K\g-\D_{K ,{\rm ext}}\rp
&= \left[ S_K,b\right]\g + \lp \g,\D_{K ,{\rm ext}}\rp
+b\D_{\chi ,{\rm ext}} \es
&= {\dot S}_K\g+b\D_{\chi ,{\rm ext}} \ ,
\ea\eqn{commslavk}

with

\eq\ba{ll}
{\dot S}_K&
= S_\chi\left[\frac{1}{8}\e_\m{}^{\n\r\l}\pa_\n
K_{\r\l}\eb\sbar^\m\right]
+{\bar S}_\chib\left[\frac{1}{8}\e_\m{}^{\n\r\l}\pa_\n
K_{\r\l}\e\s^\m\right]\es
&+S_K\left[\lx K_{\m\n}\right]
+\intx\lac 2i K_{\m\n} E^\n\dfud{}{w_\m}\rac\ .
\ea\eqn{defskdot}

Similarly,

\eq\ba{ll}
S_\k S(\g) - \BB_\g\lp S_\k\g-\D_{\k ,{\rm ext}}-\D_{\k ,{\rm gf}}\rp
&= \left[ S_\k,b\right]\g + \lp \g,\D_{\k ,{\rm ext}}\rp
+b\D_{\chi ,{\rm ext}} +\BB_\g\D_{\k ,{\rm gf}}\es
&= {\dot S}_\k\g+b\D_{\chi ,{\rm ext}}+\BB_\g\D_{\k ,{\rm gf}} \ ,
\ea\eqn{commslavkappa}

with
\eq\ba{ll}
{\dot S}_\k&
= S_\chi\left[\lp
\frac{i}{2}\pa_\m\k-\frac{1}{4}\e_\m{}^{\n\r\l}\pa_\n\k C_{\r\l}
\rp\eb\sbar^\m\right]
+{\bar S}_\chib\left[\lp
-\frac{i}{2}\pa_\m\k-\frac{1}{4}\e_\m{}^{\n\r\l}\pa_\n\k C_{\r\l}
\rp\e\s^\m\right] \es
&+S_K\left[-2\lp \pa_\m w_\n - \pa_\n w_\m\rp\right]
+S_\k\left[\lx\k\right]
+\intx\lac -2\pa_\m\k w\dfud{}{w_\m}\rac\ .
\ea\eqn{defskappadot}

And finally,

\eq\ba{lll}
\left[ S_\k , S_\chi\right]=S_\chi [-\frac{1}{2}\k\chi]\ ,&
\left[ S_\k , {\bar S}_\chib\right]
     ={\bar S}_\chib [-\frac{1}{2}\k\chib] \ ,&
\left[ S_\k , S_K\right] = S_K [2\k K_{\m\n}]\ ,\es
\left[ S_\k , \dfud{}{b^i}\right] = -2\k \dfud{}{b^i}\ ,&
\left[ S_\k , \dfud{}{w_\m}\right] = 2\k \dfud{}{w_\m}\ ,&
\left[ S_\k , \dfud{}{w}\right] = 2\k \dfud{}{w}\ ,
\ea\eqn{reglesdecomm}

complete the list of the nontrivial commutation rules of the algebra.

%%%%%%%%%%%%%%%%%%%%%%%%%%%%%%%%%%%%%%%%%%%%%%%%%%%%%%%%%%%%%
\section{New Formulation of the Model}

We have thus obtained a formulation of the supertrace identities
as  a complete algebra of symmetries,
extending the initial set of symmetries depicted in Section
\ref{sym-local} and forming a closed algebra.
Although this result looks satisfactory for the classical theory, there is
a feature which makes it unsuitable to a prompt generalization to the
quantum case. This is the presence, among the breakings
of the new symmetries,
of the terms $\D_{\k ,{\rm gf}}$ and $\D_{\chi ,{\rm gf}}$
-- originating from the gauge fixing part of the action --
 which are nonlinear in the dynamical fields (see \equ{brisurekappagf}
and \equ{supertrace4}).
Since these terms will suffer possible renormalizations, we
have to formulate a new setup better adapted to the quantum extension of
the model. This will be done by introducing a slightly modified symmetry
content, but completely equivalent to the previous one
at the classical level.

Thus, in the present section, we shall start again the construction
of the model from the very beginning, including the new symmetries
introduced in the last section in the BRS operator -- the infinitesimal
parameters $\k$, $\chi_\a$ and $K_{\m\n}$ becoming ghosts with
transformation laws assuring the nilpotency of the BRS operator.
As we
shall see, this together with the introduction of a doublet of external
fields coupled to the integrands of the nonlinear breakings solves
the ``gauge fixing problem''.

%%%%%%%%%%%%%%%%%%%%%%%%%%%%%%%%%%%%%%%%%%%%%%%%%%%%%%%%
\subsection{The $C$-gauge Problem}

With the introduction of
the shift identity, we see that $C_{\m\n}$ looses its
gauge field character since (\ref{wardk})
means that {\em any} change of $C_{\m\n}$
(and not only the changes of the form $(\pa_\m w_\n-\pa_\n w_\m)$
which correspond to a $C$-gauge transformation)
can be compensated by a suitable change of $B_\m$. This
means that the ghost fields $w_\m$ and $w$  and the $C$-gauge
transformations become useless as soon as we introduce
the field $K_{\m\n}$ and the shift operator $S_K$ (\ref{defsk}).

Thus the first tremendous change we do in the model is at the level of
the supergravitational field content, by skipping away the $C$-gauge
invariance in favor of the shift identity.
At the level of the BRS operator, care must
be taken in order to preserve
the nilpotency of $s$ on $C_{\m\n}$, which is done by imposing
the
following BRS transformation for $K_{\m\n}$:

\eq\ba{lll}
s  K_{\m\n}&=& \lx K_{\m\n}
               +2i\pa_\m\lp E_\n+C_{\n\r}E^\r\rp
               -2i\pa_\n\lp E_\m+C_{\m\r}E^\r\rp
                \ .
\ea\eqn{brsredef}

But an explicit calculation easily shows that $s$ is not
nilpotent on $K_{\m\n}$~! To get the full nilpotency, we have to
incorporate the new symmetries introduced in
the previous section in the BRS formalism. We thus promote the
fields $\k , \chi , \chib$ and $K_{\m\n}$ to the rank of ghosts and add
the
Ward operators $S_\k , S_\chi , {\bar S}_\chib$ and $S_K$ to the BRS
operator $s$ we started with in (\ref{brscomplet2}), (\ref{brscomplet3})
and (\ref{brscomplet1}).

We then construct the BRS transformations of the
new ghosts by inspecting the algebra displayed in the previous section.
For
example, the term

\eq
S_\chi \left[\lx\chi+\frac{i}{4}\l_{ab}\chi\s^{ab}+i\eta\chi
  -2\lp\gP_\m\s^\m\eb\rp\chi\right]
\eqn{termedanssx}

in the r.h.s. of (\ref{dotchi}) means that the BRS transformation of
$\chi$
looks like

\eq
s\chi = \lx\chi+\frac{i}{4}\l_{ab}\chi\s^{ab}+i\eta\chi
  -2\lp\gP_\m\s^\m\eb\rp\chi +\cdots\ .
\eqn{schi}

Doing this for the whole algebra displayed in
(\ref{commslavk}) to (\ref{reglesdecomm}) and checking for the nilpotency
leads to the following BRS transformations
for the new ghost fields:

\eq\ba{lll}
s K_{\m\n}&=& \lx K_{\m\n}
              -2\k K_{\m\n}
              +2i\pa_\m\lp E_\n+C_{\n\r}E^\r\rp
              -2i\pa_\n\lp E_\m+C_{\m\r}E^\r\rp\es &&
              +4i\lp \chib\sbar_{\m\n}\eb +\chi\s_{\m\n}\e\rp
              +4\lp\chib\eb+\chi\e\rp C_{\m\n}\es
s \chi&=& \lx\chi
          +\frac{i}{4}\l_{ab}\chi\s^{ab}
          +i\eta\chi
          +\frac{1}{2}\k\chi\es &&
          -2\lp\gP_\m\s^\m\eb\rp\chi
          -2\lp\eb\gPb_\m\rp\chib\sbar^\m
          -\lp \frac{1}{8}\lp M_\m-2N_\m\rp
+\frac{i}{2}\pa_\m\k\rp\eb\sbar^\m\es
s \chib&=& \lx\chib
          -\frac{i}{4}\l_{ab}\sbar^{ab}\chib
          -i\eta\chib
          +\frac{1}{2}\k\chib\es &&
          -2\lp\e\s^\m\gPb_\m\rp\chib
          -2\lp\e\gP_\m\rp\chi\s^\m
          +\lp \frac{1}{8}\lp M_\m-2N_\m\rp
-\frac{i}{2}\pa_\m\k\rp\e\s^\m\es
s \k&=& \lx\k
        +2\lp\chib\eb+\chi\e\rp\ ,
\ea\eqn{brsnewghosts}

where we have used the notations

\eq\ba{l}
M_\m = \e_{\m}{}^{\n\r\l}\pa_\n K_{\r\l}\es
N_\m = \e_{\m}{}^{\n\r\l}\pa_\n\k C_{\r\l}
\ .\ea\eqn{defmn}

The action of the modified BRS operator $s$ on the other fields
can be found in \ref{appendixbrsoperator}.
With this generalized BRS operator, we still have,
as in (\ref{nilpotencebrisee})

\eq
s^2 =\  \mbox{equations of motion for $\l^i$ and $\p_A$}\ ,
\eqn{nilpotencebrisee2}

where the equations of motion come from the invariant
action (\ref{actioninvariante}).

\subsection{Solution of the Gauge Fixing Problem}
\label{gaugefixingproblem}

In order to write a ST identity for the new formulation of the model, we
have to solve the problem of the gauge fixing, namely we have to give a
''suitable" definition of the two nonlinear insertions

\eq\ba{l}
\O^\m =  -2\lp\sqrt{g}\cb^i g^{\m\n}A^i_\n\rp\es
\L^\m =  s\O^\m -\lx\O^\m\ ,
\ea\eqn{omegaetlambda}

which appear in the breakings of the supertrace identities
(\ref{supertrace4})
and of the Weyl symmetry (\ref{brisurekappagf}). To do this, we
introduce
a doublet of sources
$I_\m$ and $J_\m$ with the following BRS transformations:

\eq\ba{lll}
s I_\m&=& \lx I_\m
          +J_\m\\
s J_\m&=& \lx J_\m
          -2i E^\n \pa_\n I_\m
          -2i \pa_\m E^\n I_\n\ ,
\ea\eqn{brsij}

with $s^2 I_\m = s^2 J_\m = 0\ .$
We then modify the action in

\eq\ba{lll}
\S \longrightarrow &&
\S
+ s\intx \LP I_\m\O^\m\RP\es
&=& \S
+ \intx \LP J_\m\O^\m
            +I_\m\L^\m\RP\ .
\ea\eqn{actionmodifiee}

This allows the following operational definition
(which is completely suitable to a quantum treatment of the model)
for the two insertions
$\O^\m$ and $\L^\m$:

\eq\ba{l}
\O^\m =  \dfud{\S}{J_\m}\es
\L^\m =  \dfud{\S}{I_\m}\ ,
\ea\eqn{omegaetlambda2}

where we keep the notation $\S$ for the modified action.

\subsection{Classical Identities Defining the New Formulation of the
Model}

Let us now rewrite the identities defining the theory, insisting on the
changes made with respect to  Section \ref{cidtm}.

{\bf The ST identity}

The form of the ST identity has, of course, not changed, but the
external
action did since we modified the BRS operator $s$. So, we have

\eq
S(\S)=0\ ,
\eqn{STidentity2}

with

\eq
\S = \S_{\rm inv} +\S_{\rm ext}
\hspace{6mm}\mbox{and}\hspace{6mm}
\S_{\rm inv} = \S_{\rm SYM} +\S_{\rm gf}\ ,
\eqn{nouvelleaction}

where $\S_{\rm SYM}$ is given in (\ref{actionfiltree}-\ref{SYM4}),
$\S_{\rm ext}$ is as in (\ref{actionexterieure}) but with
the BRS operator $s$ modified
by \equ{brsnewghosts} and \equ{brsij} (see \ref{appendixbrsoperator},
equations (\ref{brsgen1}-\ref{brsgen3}) for the action of $s$ on
the other fields)
and $\S_{\rm gf}$ is the sum of the gauge fixing action
of the old formulation (\ref{gfaction})
and of the modification done in (\ref{actionmodifiee}).
It can be written

\eq\ba{rl}
\S_{\rm gf} = s\intx \sqrt{g}&\lac
 -\lp \pa_\m +2I_\m\rp\cb^i g^{\m\n} A^i_\n\rac\ .
\ea\eqn{newgfaction}

{\bf The gauge condition}

\eq
\dfud{\S}{b^i(x)}=\O_b^i(x) \ ,
\eqn{gaugecondition2}

where $\O_b^i=\lp\pa_\m -2I_\m\rp\lp \sqrt{g} g^{\m\n} A^i_\n\rp$
is still a classical
breaking, but is different from the one in (\ref{gaugecondition}).

{\bf The ghost identities}

The following ghost identities are unchanged:

\eq
F^i\S=\D_{\rm G}^i\ ,\hspace{1cm}
\dfud{\S}{\xi^\m(x)}=\O_{{\rm D}\m}(x)\ ,\hspace{1cm}
\dfud{\S}{\l_{ab}(x)}=\O^{ab}_{{\rm L}}(x)\ ,\hspace{1cm}
\dfud{\S}{\eta(x)}=\O_{\rm R}(x)\ .
\eqn{ghostequations2}

Finally, the ghosts identities associated with the new fields are:

\eq
\dfud{\S}{K_{\m\n}(x)}=0\ ,\hspace{1cm}
\dfud{\S}{\chi(x)}=\O_\chi(x)\ ,\hspace{1cm}
\dfud{\S}{\chib(x)}={\bar \O}_\chib(x)\ ,\hspace{1cm}
\HH_\k (x)\S=\O_\k(x)\ ,
%\dfud{\S}{\k(x)}+\pa_\m\dfud{\S}{J_\m(x)}=\O_\k(x)\ ,
\eqn{ghostequations3}

where the classical breakings $\O_\chi(x), {\bar \O}_\chib(x)$
and $\O_\k(x)$
are given in \ref{appendixd} and where

\eq
\HH_\k (x) =  \dfud{}{\k(x)}+\pa_\m\dfud{}{J_\m(x)}\ .
\eqn{defhkappa}

Again, we can recover the different linear symmetries
(now including the shift
identity, the supertrace identities and the Weyl symmetry),
by commuting the identities
(\ref{gaugecondition2}-\ref{ghostequations3}) with the ST identity
(\ref{STidentity2}).
This is explicitly done in \ref{appendixd}.

%%%%%%%%%%%%%%%%%%%%%%%%%%%%%%%%%%%%%%%%%%
\subsection{Supercurrent Ward Identities}

\subsubsection*{Conservation Laws:}
We identify the components of the supercurrent multiplet of the new formulation
in the same way
as in the old formulation (see \ref{currents}), and we use the Ward identities
for local $R$-invariance, diffeomorphisms and supersymmetry displayed in
\ref{appendixd} to get the conservation laws. We thus find that the
conservation laws for the current $R$ and for the supersymmetry
are unchanged with respect to \equ{R-T-conserv} and \equ{Q-conserv},
whereas the conservation law for the energy-momentum
tensor becomes:

\eq\ba{ll}
\pa_\m T^\m{}_\n =  &\sumall \pa_\n\vf\dfud{\S}{\vf}
+ \pa_\m \LP\d_\n^\m\displaystyle{\sum_{\vf^*\in I^*}\vf^*\dfud{}{\vf^*}}
-A^i_\n\dfud{}{A^i_\m} + A^{*i\m}\dfud{}{A^{*i\n}}
%\es&
-\gP_\n\dfud{}{\gP_\m}
+\gPb_\n\dfud{}{\gPb_\m}
\es&
-C_{\n\r}\dfud{}{C_{\m\r}}
-B_\n\dfud{}{B_\m}
-K_{\n\r}\dfud{}{K_{\m\r}}
-I_\n\dfud{}{I_\m}
-J_\n\dfud{}{J_\m}
\RP \S
+\pa_\m\lp\xi^\m\dfud{\S}{\xi^\n}\rp
\ .
\ea\eqn{R-T-conserv2}

Again, the local Lorentz invariance (fourth of \equ{symmlin},
with \equ{omegalorentz}) shows that the energy-momentum tensor is symmetric:

\eq\ba{rl}
T^{[ab]}=&
\frac{i}{4}\l^i\s^{ab}\dfud{\S}{\l^i}
-\frac{i}{4}\lb^i\s^{ab}\dfud{\S}{\lb^i}
+\frac{i}{4}\p_A\s^{ab}\dfud{\S}{\p_A}
-\frac{i}{4}\pb_A\s^{ab}\dfud{\S}{\pb_A}
+\frac{i}{4}\gP_\m\s^{ab}\dfud{\S}{\gP_\m}
\es &
-\frac{i}{4}\gPb_\m\s^{ab}\dfud{\S}{\gPb_\m}
+\frac{i}{4}\e\s^{ab}\dfud{\S}{\e}
-\frac{i}{4}\eb\s^{ab}\dfud{\S}{\eb}
+          \lp\d^a_c\l^b{}_d
              -\d^b_c\l^a{}_d\rp\dfud{\S}{\l_{cd}}
\es &
+\frac{i}{4}\chi\s^{ab}\dfud{\S}{\chi}
-\frac{i}{4}\chib\s^{ab}\dfud{\S}{\chib}
+\pa_\m\lp\frac{1}{2}\xi^\m\dfud{\S}{\l_{ab}}\rp\ .
\ea\eqn{improvedtmn2}

%%%%%%%%%%%%%%%%%%%%%%%%%%%%%%%%%%%%%%
\subsubsection*{Supertrace Identities:}
We proceed in the same way with the trace identities
(second of (\ref{symlinnew}), with (\ref{omegasupertrace})
and last of (\ref{symlinnew}), with (\ref{omegaweyl}))
corresponding to the supertrace identity and to the Weyl identity,
respectively, to get:

\eq\ba{ll}
\sbar_\m Q^\m &=
 -2\fb_a\dfud{\S}{\pb_a}
 -2\pb^*_a\dfud{\S}{\fb^*_a}
 +3\sbar^\n\s_\m\gPb_\n\dfud{\S}{B_\m}
 +i\sbar_{ab}\eb\dfud{\S}{\l_{ab}}
 +4\eb\cb^i\dfud{\S}{b^i}%\es&
 +3i\eb\dfud{\S}{\eta}\es&
 -2\eb\dfud{\S}{\k}
 +2\lp\eb\gPb_\m\rp\sbar^\m\dfud{\S}{\chi}
 +\pa_\m\lp\xi^\m\dfud{\S}{\chib}\rp
 +\lp-\frac{i}{4}\l_{ab}\sbar^{ab}
          +i\eta
          -\frac{1}{2}\k
          +2\lp\e\s^\m\gPb_\m\rp\rp\dfud{\S}{\chib}
\ ,\es
T^\m{}_\m &=
\displaystyle{
\sum_{\parbox{13mm}{\tiny{all fields \\ except $\vb{\m}{a}$}}}}
{\widetilde d}_\vf\vf\dfud{\S}{\vf}
-\frac{1}{4}\pa_\l\lac
\e_\r {}^{\l\m\n}C_{\m\n}\lp
3\dfud{\S}{B_\r}
+\eb\sbar^\r\dfud{\S}{\chi}
+\e\s^\r\dfud{\S}{\chib}
\rp\rac \es &
+\pa_\m\lac
\frac{i}{2}\eb\sbar^\m\dfud{\S}{\chi}
-\frac{i}{2}\e\s^\m\dfud{\S}{\chib}
+\dfud{\S}{I_\m}
+\xi^\m\lp\dfud{\S}{\k}+\pa_\n\dfud{\S}{J_\n}\rp
\rac
\ .
\ea\eqn{supertraceids}

The second of these identities shows that $T^\m{}_\n$ is indeed
the improved energy-momentum tensor, being on-shell traceless.

Finally, we explicitly give the shift identity
(first of (\ref{symlinnew}), with (\ref{omegashift}))

\eq
\dfud{\S}{C_{\m\n}}=
-\frac{1}{4}\pa_\l\lac
\e_\r {}^{\l\m\n}C_{\m\n}\lp
3i\dfud{\S}{B_\r}
+\eb\sbar^\r\dfud{\S}{\chi}
+\e\s^\r\dfud{\S}{\chib}
\rp\rac
\ ,
\eqn{shiftequ}

which shows that the insertion $\ \dfud{\S}{C_{\m\n}}\ $ is not independent.

%%%%%%%%%%%%%%%%%%%%%%%%%%%%%%%%%%%%%%
\subsection{Flat Limit}\label{lim-plate-conf}

Defining the F.L. of the new formulation of the model will allow to
 recognize in more familiar terms the
 improvements implied by the  adjunction of
the new symmetries. First, to keep
 track of the supertrace  identity
and of the Weyl transformations when going to the F.L., we define

\eq\ba{lll}
\chi&\stackrel{\rm F.L.}{\longrightarrow}&\fl{\chi}\es
\k&\stackrel{\rm F.L.}{\longrightarrow}&\fl{\k}\ .
\ea\eqn{flghosts2}

Next we study the constraint of coherence (\ref{flvf}), which leads to

\eq\ba{lll}
K_{\m\n}&\stackrel{\rm F.L.}{\longrightarrow}&0\es
\e&\stackrel{\rm F.L.}{\longrightarrow}&\fl{\e}-i x^\m\fl{\chib\sbar_\m}\ ,
\ea\eqn{flghosts3}

the first  one coming from imposing $\ \fl{sC_{\m\n}}=0\ $ and the
second from imposing
$\ \fl{s\gP_\m}=0\ $, since both $C_{\m\n}$ and $\gP_\m$
have a null F.L. (See Section \ref{lim-plate-poinc}).

Applying then (\ref{flvf}) to $\k$ leads to a new problem: the r.h.s tends
to an $x$-dependent term
(coming from the $x$-dependent term of the flat
limit of $\e$) whereas the l.h.s is constant. This problem is solved by
introducing a new constant ghost $\fl{\b_\m}$ and to redefine
the F.L. of $\k$ as

\eq
\k\stackrel{\rm F.L.}{\longrightarrow}\fl{\k}+x^\m\fl{\b_\m}\ .
\eqn{flghosts5}

Finally applying (\ref{flvf}) to the remaining ghosts leads to

\eq\ba{lll}
\xi^\m&\stackrel{\rm F.L.}{\longrightarrow}&
\fl{\xi^\m}
-\fl{\l^\m {}_a}x^a
+\fl{\k}x^\m
-\frac{1}{2}x^2\fl{\b^\m}
+x^\m x^\n\fl{\b_\n}
\es
\l_{ab}&\stackrel{\rm F.L.}{\longrightarrow}&
\fl{\l_{ab}}
+\fl{\b_a} x_b
-\fl{\b_b} x_a
\es
\e&\stackrel{\rm F.L.}{\longrightarrow}&\fl{\e}-i x^\m\fl{\chib\sbar_\m}\es
\eta&\stackrel{\rm F.L.}{\longrightarrow}&\fl{\eta}\es
K_{\m\n}&\stackrel{\rm F.L.}{\longrightarrow}&0\es
\chi&\stackrel{\rm F.L.}{\longrightarrow}&\fl{\chi}\es
\k&\stackrel{\rm F.L.}{\longrightarrow}&\fl{\k}+x^\m\fl{\b_\m}\ .
\ea\eqn{flghosts4}

Looking at the F.L. of the transformations laws of the dynamical fields and
at the algebra encoded in (\ref{flghosts4})
clearly shows that we have thus recovered the whole superconformal
algebra~\cite{frad-tsey,ps-book},
where the ghosts $\fl{\k},\fl{\b_\m}$ and $\fl{\chi}$
are the ghosts for the dilatations, the special conformal transformations
and the  special superconformal transformations respectively.

\section{Stability of the solution}

As a first step to a complete quantum treatment of the model, let us
study
the stability of the classical action $\S$ (\ref{nouvelleaction}) under
radiative corrections. That means looking for the most general
classical integrated insertion $\S_c$ (having the same quantum numbers 
and dimension as $\S$) such that the perturbed action

\eq
\S' = \S + \zeta\S_c\ ,
\eqn{pertaction}

satisfies, at the first order in the infinitesimal
parameter $\zeta$, the same constraints as $\S$, \ie the ST identity
(\ref{STidentity2}), the gauge condition  (\ref{gaugecondition2})
and the ghost identities (\ref{ghostequations2}) and
(\ref{ghostequations3}).
The action is said to be stable if  $\S_c$ corresponds
to genuine counterterms, \ie
if  $\S_c$ can be reabsorbed through a redefinition of the  field
amplitudes and of the parameters (namely the coupling
constants).

Expanding these constraints on $\S'$ at the first order in $\zeta$ leads
to
the following set of equations that must be satisfied by $\S_c$:

\eq
\BB_\S\S_c=0\ ,
\eqn{contrbrs}
\eq\ba{l}
\dfud{\S_c}{b^i}=0\ ,\hspace{1cm}
F^i\S_c =0\ ,\es
\dfud{\S_c}{\xi^\m}=\, \dfud{\S_c}{\l_{ab}}=\,\dfud{\S_c}{\eta}=0\ ,\es
\dfud{\S_c}{K_{\m\n}}=\, \dfud{\S_c}{\chi}=\,\dfud{\S_c}{\chib}=\,
\HH_\k\S_c =0\ .
\ea\eqn{contrgheq}

The constraints (\ref{contrgheq}) are easily solved and lead to a $\S_c$
which is independent of $b^i,\xi^\m,\l_{ab},\eta,K_{\m\n},\chi,\chib$
and which depends on $c^i$ only through its derivative $\pa_\m c^i$
and on $\k$ only through the combination

\eq
{\hat J}_\m = J_\m +\pa_\m \k\ .
\eqn{defhatj}

The equations (\ref{contrbrs}) constitutes a cohomology problem, due to
the
nilpotency of the linearized ST operator $\BB_\S$
(see (\ref{nilpotencedeSTlinearise})). The strategy we applied to
construct
the explicit solution is given in \ref{contretermes} and we give here
the result in the form:

\eq
\S_c = \S_{\rm ph} +\BB_\S {\hat \S}_c\ ,
\eqn{contrt}

with

\eq
\S_{\rm ph}=Z_G \dpad{\S}{G}
+Z_{(ABC)}\dpad{\S}{\L_{(ABC)}}
+{\bar Z}_{(ABC)}\dpad{\S}{\Lb_{(ABC)}}
+\S_{\rm CSG}
\ ,
\eqn{explicites1}

and

\eq\ba{rl}
{\hat \S}_c=\intx \LP &\!\!\!
    \displaystyle{
Z_A A^i_\m {\hat A}^{*i\m}
+Z_\l \l^{*i}\l^i
+{\bar Z}_\lb \lb^{*i}\lb^i
                  }
\\
&\!\!\!
    \displaystyle{
+Z_{\f AB} \lp \f^*_A +\frac{1}{2}\gPb_\m\sbar^\m\p^*_A \rp\f_B
+{\bar Z}_{\f AB}\lp \fb^*_A -\frac{1}{2}\gP_\m\s^\m\pb^*_A\rp\fb_B
                  }
\\
&\!\!\!
    \displaystyle{
+Z_{\p AB}\p^*_A\lp \p_B +\frac{1}{2}\s^\m \gPb_\m\f_B\rp
+{\bar Z}_{\p AB}\pb^*_A\lp \pb_B-\frac{1}{2}\sbar^\m \gP_\m\fb_B\rp
                  }
\\
&\!\!\!
    \displaystyle{
+V g^{\m\r}\lp \gP_\m +\frac{i}{3}\gP_\n \s^\n{}_\m\rp\l^{*i} A^i_\r
+{\bar V} g^{\m\r}\lp \gPb_\m +\frac{i}{3}\gPb_\n \sbar^\n{}_\m\rp
                       \lb^{*i} A^i_\r
                  }
\RP\ .
\ea\eqn{explicites2}

The constant $Z_G$ and the $\GG$-invariant tensor $Z_{(ABC)}$
are interpreted as renormalizations of the coupling constants
$G$ and $\L_{(ABC)}$ respectively, whereas the constants
$Z_A,Z_\l,V$ and the $\GG$-invariant tensors $Z_{\f AB},Z_{\p AB}$
correspond to unphysical field renormalizations. Finally, the
counterterm $\S_{\rm CSG}$ -- which is independent of the dynamical fields --
is in fact the conformal supergravity action~\cite{csg,frad-tsey},
but is interpreted in our context of
external supergravity as corrections to the current algebra.
The complete form of $\S_{\rm CSG}$ can be found in~\cite{csg,frad-tsey},
and it starts as

\eq
\S_{\rm CSG}=Z_{\rm CSG}\intx\sqrt{g}\lp
C^{\m\n\r\l} C_{\m\n\r\l}
-\frac{3}{4} {\hat G}^{\m\n} {\hat G}_{\m\n}
+\cdots
\rp
\ ,
\eqn{csg02}

where the Weyl tensor $C^{\m\n\r\l}$ and the $R$-curvature tensor
${\hat G}_{\m\n}$ are defined in \ref{contretermes}, equations
\equ{weyltensor} and \equ{rcurvtensor}.

\underline{Remark}: At the F.L. the $V$-term and $\S_{\rm CSG}$
disappear and the other ones
tend, as expected, to the counterterms found in the rigid
model~\cite{I}.

\section{Conclusion}
%*************************
 We have thus succeeded in constructing a general functional
formalism which incorporates, within a unique Slavnov-Taylor identity
associated to a general BRS invariance, both the set of
conservation laws of the supercurrent components, and the set of the
``supertrace identities''. The latter includes the --
potentially anomalous -- Ward identities for the traces of the
energy-momentum tensor and spinor current \equ{supertraceids},
as well as the shift identity \equ{shiftequ}.
In order to achieve this, we had to reformulate the ``new minimal''
external supergravity in such a way that
the conformal properties
implied by the supertrace identities might be
naturally incorporated.

The supermultiplet structure of the supertrace identities -- which is
obvious in the superspace approach, where they form a superfield -- is
here somewhat hidden in the algebra of the Slavnov-Taylor operator. One can
nevertheless observe it in the algebraic identities
\equ{commslavchi},\equ{commslavk},\equ{commslavkappa}
and \equ{reglesdecomm}.
We have given the algebra explicitly for the old formulation.
But, in the new formulation, designed indeed in order to take care of it,
the complete algebra can be read out directly from
the BRS transformations of the ghosts (see \equ{brsnewghosts}).

%%%%%%%%%%%%%%%%%%%%%%%%%%%%%%%%%%%%%%%%%%%%%%%%%%%%%%%%%%%%%%%
\noindent{\bf Acknowledgments.\,} We thank Nicola Maggiore and
Klaus Sibold for interesting and useful discussions at a
preliminary stage of this work.
One of us (S.W.) is indebted to the fund Fanny Wurth, University
of Geneva, Switzerland, for its financial support,
and thanks the  Physics Department of the Universidade Federal do
Esp\'\i rito Santo, Brazil, for its kind hospitality.

%*********Appendix********
\renewcommand{\theequation}{\Alph{section}.\arabic{equation}}
\renewcommand{\thesection}{Appendix \Alph{section}}
%*************************
%*************************

\setcounter{section}{0}

\section{Quantum Numbers of the Fields}
\label{appendicetableau}

We give in this Appendix
the ghost numbers, Grassmann parities, dimensions and $R$-weights
of all the fields appearing in the model.

They commute or anticommute according to the
formula:

\eq
\vf_1 \vf_2 = {(-1)}^{{\rm GP}(\vf_1) \cdot {\rm GP}(\vf_2)} \vf_2 \vf_1
\ .
\eqn{reglesdecommutation}
%%%%%%%%%%%%%%%%%%%%%%%%%%%%%%%%%%%%%%%%%%%%%%%%%%%
\begin{table}[ht]%[hbt]
\centering
\begin{tabular}{|c||c|c|c|c|c|c|c|c|c|c|c|c||c|c|c|c|}
\hline
&$A^i_\mu$
&$\l^i$
&$\f_A$
&$\p_A$
&$c^i$
&$\cb^i$
&$b^i$
&$A^{*i\m}$
&$\l^{*i}$
&$\f^*_A$
&$\p^*_A$
&$c^{*i}$
&$\vb{a}{\m}$
&$\gP_\m$
&$B_{\m}$
&$C_{\m\n}$
\\
\hline\hline
$GP$&0&1&0&1&1&1&0&1&0&1&0&0&0&1&0&0\\
\hline
$\F\Pi$&0&0&0&0&1&-1&0&-1&-1&-1&-1&-2&0&0&0&0\\
\hline
$d$&1&3/2&1&3/2&0&2&2&3&5/2&3&5/2&4&0&1/2&1&0\\
\hline
$R$&0&-1&-2/3&1/3&0&0&0&0&1&2/3&-1/3&0&0&-1&0&0 \\
\hline
\end{tabular}
%\caption[t1]\hfill{\small\parbox{15cm}{ \ \\
%Grassmann parity $GP$, ghost number $\F\Pi$, dimensions  $d$,
%  and $R$-weight of the dynamical fields and their antifields.}}
%\label{tabledyn}
\end{table}

\begin{table}[ht]%[hbt]
\centering
\begin{tabular}{|c||c|c|c|c||c|c||c|c|c||c|c|}
\hline
&$\xi^\m$
&$\l_{ab}$
&$\e$
&$\eta$
&$w_\m$
&$w$
&$K_{\m\n}$
&$\chi$
&$\k$
&$I_\m$
&$J_\m$
\\
\hline\hline
$GP$&1&1&0&1&1&0&1&0&1&0&1\\
\hline
$\F\Pi$&1&1&1&1&1&2&1&1&1&0&1\\
\hline
$d$&-1&0&-1/2&0&-1&-2&0&1/2&0&1&1\\
\hline
$R$&0&0&-1&0&0&0&0&1&0&0&0 \\
\hline
\end{tabular}
%\label{tableextghost}
\end{table}
 The tables above show the Grassmann parities $GP$,
the ghost numbers $\F\Pi$, the dimensions  $d$
  and the $R$-weights of the various fields.

%%%%%%%%%%%%%%%%%%%%%%%%%%%%%%%%%%%%%%%%%%%%%%%%%%%%%%
\section{Notations and Conventions}
\label{appendixa}

\point{Units:} $\hbar=c=1$.
\point{Flat space-time metric:} $(\eta_{ab}) = \mbox{diag}(1,-1,-1,-1)
    \ ,\quad(a,b=0,1,2,3)$.
%\point{Fourier transform:}
%\[
%f(x) = \dfrac{1}{2\pi}\int dk\;e^{ikx}\tilde f(k)\ ,\quad
%\tilde f(k) = \int dx\;e^{-ikx}f(k)\ .
%\]
\point{Weyl spinor:} $(\p_\a\ ,\ \a =1,2)\
   \in\, \mbox{repr. }(\half,0)$ of the Lorentz group.
%  The spinor components are Grassmann variables:
$\p_\a\p'_\b=-\p'_\b\p_\a$
\point{Complex conjugate spinor:}
    $(\bar\p_\da\ ,\ \da =1,2)\
                           \in\, \mbox{repr. }(0,\half)$.
\point{Raising and lowering of spinor indices:}
  \[\ba{l}
  \p^\a=\e^{\a\b}\p_\b\ ,\quad \p_\a=\e_{\a\b}\p^\b\ ,\es
  \mbox{with }   \e^{\a\b}=-\e^{\b\a}\ ,\quad\e^{12}=1\ ,\quad
  \e_{\a\b}=-\e^{a\b}\ ,\quad \e^{\a\b}\e_{\b\g}=\d^\a_\g\ ,\es
  \mbox{(the same for dotted indices).}
  \ea\]
\point{Derivative with respect to a spinor component:}
  \[\ba{l}
  \dpad{}{\p^\a}\p^\b=\d^{\b}_{\a}\ ,\quad
     \dpad{}{\p_\a} = \e^{\a\b}\dpad{}{\p^\b}\ ,\es
  \mbox{(the same for dotted indices).}
  \ea\]
\point{Pauli matrices:}
  \[\ba{l}
  \lp\s^a_{\a\db}\rp=
  \lp\, \s^0_{\a\db},\, \s^1_{\a\db},\,
              \s^2_{\a\db},\, \s^3_{\a\db}\, \rp\ , \es
  \sbar_a^{\da\b}=\s_a^{\b\da}
    =\e^{\b\a}\e^{\da\db}\s_{a\,\a\db}\ ,\es
   {(\s^{ab})_\a}^\b=
    \dfrac{i}{2}{\lc \s^a\sbar^b-\s^b\sbar^a \rc_\a}^\b   \ , \quad
  {(\sbar^{ab})^\da}_\db=
    \dfrac{i}{2}{\lc \sbar^a\s^b-\sbar^b\s^a \rc^\da}_\db   \ ,
\ea\]
with
  \[\ba{l}
  \s^0=\lp\matrix{1&0\\0&1}\rp\ ,\quad  \s^1=\lp\matrix{0&1\\1&0}\rp\
,\quad
    \s^2=\lp\matrix{0&-i\\i&0}\rp\ ,\quad
                \s^3=\lp\matrix{1&0\\0&-1}\rp\ ,\\[5mm]
  \sbar^0 =\s^0\ ,\quad \sbar^i=-\s^i=\s_i\ ,\quad
  \s^{0i} = -\sbar^{0i} = -i\s^i\ ,\quad
   \s^{ij} = \sbar^{ij} = \e^{ijk}\s^k\ ,\es
   i,j,k=1,2,3\ .
  \ea\]
\underline{Remark}: We  define $\s_\m \equiv \vb{\m}{a}\s_a$.  (See
\ref{appendixb} for the definition of the vierbein $\vb{\m}{a}$.)
\point{Summation conventions and complex conjugation:}
Let $\p$ and $\chi$ be two Weyl (anticommutant) spinors; we have:
  \[\ba{l}
  \p\chi=\p^\a\chi_\a = -\chi_\a\p^\a = \chi^\a\p_\a = \chi\p \ ,\es
  \pb\chib=\pb_\da\chib^\da = -\chib^\da\pb_\da
        = \chib_\da\pb^\da  = \chib\pb \ ,\es
  \p\s^\m\chib = \p^\a\s^\m_{\a\da}\chib^\da\ ,\quad
        \pb\sbar_\m\chi = \pb_\da\sbar_\m^{\da \a}\chi_\a\ ,\es
  \overline{(\p\chi)} = \chib\pb = \pb\chib \ ,\es
  \overline{(\p\s^\m\chib)} = \chi\s^\m\pb = -\pb\sbar^\m\chi\ ,\es
  \overline{(\p\s^{\m\n}\chi)} = \chib\sbar^{\m\n}\pb\ .
\ea\]
\point{$\e$-tensor:}
$\e_{\m\n\r\l}\equiv\e_{abcd}\vb{\m}{a}\vb{\n}{b}\vb{\r}{c}\vb{\l}{d}\
,$
\hspace{5mm}$\e_{0123} = +1\ .$
\point{Lie derivative:}
The Lie derivative $\lx$ along
$\xi^\m$ is defined,  for a scalar or a vector field, by:

\[\ba{lll}
\lx\vf &=& \xi^\m\pa_\m\vf \es
\lx\vf_\n &=& \xi^\m\pa_\m\vf_\n +\pa_\n\xi^\m \vf_\m\es
\lx\vf^\n &=& \xi^\m\pa_\m\vf^\n -\pa_\m\xi^\n \vf^\m\ ,
\ea\]

with the obvious generalization to the higher rank tensors.  For
a tensorial density $\Phi^{\cdots}$, the formula

\[
\lx\Phi^{\cdots} = e\lx \lp e^{-1}\Phi^{\cdots}\rp
  + \pa_\m (\xi^\m e) e^{-1}\Phi^{\cdots}\ ,
\]

defines its Lie derivative from the Lie derivative of the tensor
$e^{-1}\Phi^{\cdots}$.

\section{Vierbein Formalism}
\label{appendixb}

As already said, we suppose that the space-time torsion
is identically zero, which means
that the Lorentz connection $w_{\m ab}$  is expressed as
a function of the vierbein $\vb{\m}{a}$:

\eq
w_{\m ab}=\frac{1}{2}\lp
          \vbb{\n}{a}\pa_\m \vb{b}{\n}
          -\vbb{\n}{b}\pa_\m \vb{a}{\n}
          -\vb{a}{\n}\vb{b}{\r}
          (\pa_\n g_{\r\m} - \pa_\r g_{\n\m})\rp\ .
\eqn{lorentzconnection}

The indices $\m,\n,\cdots=0,\cdots,3$ and  $a,b,\cdots=0,\cdots,3$ are
world and tangent
space indices, respectively.

The inverse vierbein is given by $\vb{a}{\m}$, with:

\eq
\vb{a}{\m}\vb{\m}{b}=\d_a^b\ ,\hspace{3cm}
\vb{\m}{a}\vb{a}{\n}=\d_\m^\n\ .
\eqn{vierbein}

The metric is then given by:

\eq
g_{\m\n} = \vb{\m}{a}\vbb{\n}{a}\ .
\eqn{metric}

The metric $g_{\m\n}$ is used to raise and lower world indices,
the flat metric $\eta_{ab}$ is used to raise and lower the tangent
space indices and
the vierbein $\vb{\m}{a}$ is used to transform a tangent space indice
into a world indice and vice versa:

\eq\ba{llllll}
g_{\m\n}\vf^\n &=&\vf_\m \ ,\hspace{3cm}&
g^{\m\n}\vf_\n &=&\vf^\m\ ,\es
\eta_{ab}\vf^b &=&\vf_a\ , &
\eta^{ab}\vf_b &=&\vf^a\ ,\es
\vb{\m}{a}\vf_a &=&\vf_\m\ ,&
\vb{a}{\m}\vf_\m &=&\vf_a\ ,
\ea\eqn{monteeetdescente}

where $g^{\m\n}$ is the inverse metric.

The Christoffel symbols are defined by:

\eq
\G_\m {}^\l {}_\n =
          \frac{1}{2}g^{\l\r}\lp
          \pa_\r g_{\m\n}
          -\pa_\m g_{\r\n}
          -\pa_\n g_{\m\r}\rp\ .
\eqn{christoffelsymbols}

And finally, the Riemann tensor, the Ricci tensor and
the scalar curvature are  defined by:

\eq\ba{rl}
R_{\m\n ab} =& \pa_\n w_{\n ab}
              -\pa_\m w_{\m ab}
              +w_{\n}{}_a{}^c w_{\m cb}
              -w_{\m a}{}^c w_{\n cb}\ ,\es
R_{\m\n}=&R_{\m\l ab}\vbh{b}{\l}\vb{\n}{a}\ ,
\es
R=&g^{\m\n}R_{\m\n}\ .
\ea\eqn{riemann}

\section{Covariant Derivative}
\label{appendixc}

A generic field $\vf$ is completely characterized by

\eq\ba{ll}
\bullet&
\mbox{its world indices:}\ \m ,\n , \r ,\ldots
\es \bullet&
\mbox{its Lorentz indices:}\left\{\ba{ll}
  a,b,c,\ldots &\mbox{tangent space vectorial indices}\es
  \a ,\b ,\g ,\ldots &\mbox{undotted spinorial indices}\es
  \da ,\db ,\dg ,\ldots &\mbox{dotted spinorial indices}\es
  \ea\right.
\es \bullet&
\mbox{its $\GG$-gauge indices:}\left\{\ba{ll}
  i,j,k,\ldots &\mbox{of the adjoint representation of $\GG$}\es
  A,B,C,\ldots &\mbox{of some rep. $R$ of $\GG$}
  \ea\right.
\es \bullet&
\mbox{its $R$-weight:}\ R_\vf
\es \bullet&
\mbox{its canonical dimension:}\ d_\vf\ .
\ea\eqn{defchamp}

The covariant derivative $\nabla_\m\vf$  of a tensor $\vf$
is then defined by

\eq\ba{lll}
\nabla_\m\vf=&\pa_\m\vf&\es&
 +\G_\m {}^\l {}_\n\vf_\l
 &\mbox{for each covariant indice}\ \n\ \mbox{of}\ \vf_\n \es&
 -\G_\m {}^\n {}_\l\vf^\l
 &\mbox{for each contravariant indice}\ \n\ \mbox{of}\ \vf^\n \es&
  +w_{\m a} {}^{b}\vf_b
 & \mbox{for each tangent space Lorentz indice}\ a\ \mbox{of} \ \vf_a
\es&
  +\frac{i}{4}w_{\m ab} \vf^\b \s^{ab}{}_{\b}{}^\a
 & \mbox{for each undotted Lorentz indice}\ \a\ \mbox{of} \ \vf^\a \es&
  -\frac{i}{4}w_{\m ab} \sbar^{ab}{}^\db {}_\dg\vf^\dg
 & \mbox{for each dotted Lorentz indice}\ \db\ \mbox{of} \ \vf^\db \es&
 +f^{ijk}A^j_\m\vf^k
 &\mbox{for each gauge indice}\ i\ \mbox{of} \ \vf^i \es&
  +T^i_{AB}A^i_\m\vf_B
 & \mbox{for each gauge indice}\ A\ \mbox{of} \ \vf_A \es&
  -iR_\vf B_\m \vf &
\ea\eqn{covder}

\section{BRS Operator in the New Formulation}
\label{appendixbrsoperator}

The BRS operator $s$ including all the symmetries of the new formulation
of
the model is given by:

\eq\ba{lll}
s  A^{i}_\m &=& \lx A_\m^i
                + \nabla_\m c^i
                +\e\s_\m\lb^i
                +\l^i\s_\m\eb\\
s  \l^i   &=& \lx\l^i
               -f^{ijk} c^j \l^k
               +\frac{i}{4}\l_{ab}\l^i\s^{ab}
               -i\eta\l^i
               %+\frac{3}{2}\kzero\l^i
               -\frac{1}{2}\e\s^{\m\n}{\widetilde F}^i_{\m\n}
               -\frac{i}{2}G^2(\fb_AT^i_{AB}\f_B)\e
               +\frac{3}{2}\k\l^i\\
s  \lb^i &=&  \lx\lb^i
               -f^{ijk} c^j \lb^k
               -\frac{i}{4}\l_{ab}{\bar\s}^{ab}\lb^i
               +i\eta\lb^{i\db}
               %+\frac{3}{2}\kzero\lb^i
               +\frac{1}{2}{\bar \s}^{\m\n}{}\eb {\widetilde F}^i_{\m\n}
               +\frac{i}{2}G^2(\fb_AT^i_{AB}\f_B)\eb
               +\frac{3}{2}\k\lb^i\\
s \f_A &=& \lx\f_A
            -T^i_{AB}c^i\f_B
            -\frac{2}{3}i\eta\f_A
             +2\e\p_A
            +\k\f_A\\
s \fb_A &=& \lx\fb_A
             -T^i_{AB}c^i\fb_B
             +\frac{2}{3}i\eta\fb_A
             -2\eb\pb_A
             +\k\fb_A\\
s \p_A &=& \lx\p_A
            -T^i_{AB}c^i\p_B
            +\frac{i}{4}\l_{ab}\p_A\s^{ab}
            +\frac{1}{3}i\eta\p_A
            %+\frac{3}{2}\kzero\p_a
            +i\eb\sbar^\m{\cal D}_\m\f_A
            +2\Lb_{(ABC)}\fb_B\fb_C\e
            +\frac{3}{2}\k\p_A
            +2\f_A\chi\\
s \pb_A &=& \lx\pb_A
             -T^i_{AB}c^i\pb_B
             -\frac{i}{4}\l_{ab}{\bar\s}^{ab}\pb_A
             -\frac{1}{3}i\eta\pb_A
             %+\frac{3}{2}\kzero\pb_a
             +i\sbar^\m\e{\cal D}_\m\fb_A
             -2 \L_{(ABC)}\f_B\f_C\eb
             +\frac{3}{2}\k\pb_A
             -2\fb_A\chib\\
s  c^i &=& \lx c^i
           -\frac{1}{2}f^{ijk}c^jc^k
           -2iE^\m A^i_\m\\
s \cb^i &=& \lx\cb^i
             + b^i
             +2\k\cb^i
             \\
s  b^i &=& \lx b^i
            -2i E^\m\pa_\m\cb^i
            +2\k b^i
            -4(\chib\eb+\chi\e)\cb^i
\ea\eqn{brsgen1}
\eq\ba{lll}
s \vb{\m}{a}&=& \lx\vb{\m}{a}
                 +\l^a {}_b\vb{\m}{b}
                 %-\kzero\vb{\m}{a}
                 +2\e\s^a\gPb_\m
                 +2\gP_\m\s^a\eb
                 -\k\vb{\m}{a}\\
s \gP_\m&=& \lx\gP_\m
             +\frac{i}{4}\l_{ab}\gP_\m\s^{ab}
             -i\eta \gP_\m
             %-\frac{1}{2}\kzero\gP_\m
             -i{\cal D}_\m \e
             -\frac{1}{2}\k\gP_\m
             +\chib\sbar_\m\\
s \gPb_\m&=& \lx\gPb_\m
              -\frac{i}{4}\l_{ab}{\bar\s}^{ab}\gPb_\m
              +i\eta \gPb_\m
              %-\frac{1}{2}\kzero\gPb_\m
              -i{\cal D}_\m \eb
              -\frac{1}{2}\k\gPb_\m
              +\chi\s_\m\\
s  C_{\m\n}&=& \lx C_{\m\n}
               +K_{\m\n}
               -2\k C_{\m\n}
               \\ &&
               -2\e\s_\m\gPb_\n
               -2\gP_\n\s_\m\eb
               +2\e\s_\n\gPb_\m
               +2\gP_\m\s_\n\eb  \\
s  B_\m&=& \lx B_\m
           +\pa_\m \eta
           -\e\s_\m\sbar^{\n\r}{\cal D}_\n\gPb_\r
           -{\cal D}_\n\gP_\r\s^{\n\r}\s_\m\eb \\ &&
           -3\lp \chib\sbar^\n\s_\m\gPb_\n\rp
           +3\lp \chi\s^\n\sbar_\m\gP_\n\rp
           -\frac{3}{8}\lp M_\m-2N_\m\rp
\ea\eqn{brsgen2}
\eq\ba{lll}
s I_\m&=& \lx I_\m
          +J_\m\\
s J_\m&=& \lx J_\m
          -2i E^\n \pa_\n I_\m
          -2i \pa_\m E^\n I_\n
\ea\eqn{brsij2}
\eq\ba{lll}
s \xi^\m&=&  \xi^\l \pa_\l \xi^\m
             +2iE^\m\\
s \l_{ab}&=& \lx\l_{ab}
             +(\l^2)_{ab}
             +2iE^\l {\widetilde w}_{\l ab}
             -\frac{i}{2} \e_{ab\m\n} E^\m V^\n %\\ &&
             -2i\lp\chib\sbar_{ab}\eb\rp
             -2i\lp\chi\s_{ab}\e\rp\\
s \e&=& \lx \e
        +\frac{i}{4}\l_{ab}\e\s^{ab}
        -i\eta \e
        %-\frac{1}{2}\kzero\e
        +2E^\l\gP_\l
        -\frac{1}{2}\k\e\\
s \eb&=& \lx \eb
         -\frac{i}{4}\l_{ab}{\bar\s}^{ab}\eb
         +i\eta \eb
         %-\frac{1}{2}\kzero\eb
         +2E^\l\gPb_\l
         -\frac{1}{2}\k\eb\\
s \eta&=& \lx \eta
          -2iE^\l B_\l
          -3i\lp\chib\eb-\chi\e\rp\\
s K_{\m\n}&=& \lx K_{\m\n}
              -2\k K_{\m\n}
              +2i\pa_\m\lp E_\n+C_{\n\r}E^\r\rp
              -2i\pa_\n\lp E_\m+C_{\m\r}E^\r\rp\\ &&
              +4i\lp \chib\sbar_{\m\n}\eb +\chi\s_{\m\n}\e\rp
              +4\lp\chib\eb+\chi\e\rp C_{\m\n}\\
s \chi&=& \lx\chi
          +\frac{i}{4}\l_{ab}\chi\s^{ab}
          +i\eta\chi
          +\frac{1}{2}\k\chi\\ &&
          -2\lp\gP_\m\s^\m\eb\rp\chi
          -2\lp\eb\gPb_\m\rp\chib\sbar^\m
          -\lp \frac{1}{8}\lp M_\m-2N_\m\rp
+\frac{i}{2}\pa_\m\k\rp\eb\sbar^\m\\
s \chib&=& \lx\chib
          -\frac{i}{4}\l_{ab}\sbar^{ab}\chib
          -i\eta\chib
          +\frac{1}{2}\k\chib\\ &&
          -2\lp\e\s^\m\gPb_\m\rp\chib
          -2\lp\e\gP_\m\rp\chi\s^\m
          +\lp \frac{1}{8}\lp M_\m-2N_\m\rp
-\frac{i}{2}\pa_\m\k\rp\e\s^\m\\
s \k&=& \lx\k
        +2\lp\chib\eb+\chi\e\rp\ ,
\ea\eqn{brsgen3}

where we have used the notations
(\ref{dercovsupersym},\,\ref{emuetwtilde}) and \equ{defmn}.

\section{Classical Breakings, Algebra and Ward Operators for the Linear
Symmetries}\label{appendixd}

We give in this appendix the classical breakings of the Ward identities
defining the model,
as well as the algebra formed by the ST operator and the Ward
operators, which leads to the
Ward operators for $\GG$-rigid transformations, diffeomorphisms,
Lorentz transformations, $R$- and $C$-transformations and,
for the new formulation of the model,
to the Ward operators associated with the shift
identity, the supertrace identities and the
Weyl identity.
The distinction between old and new formulation is made whenever it is
necessary.

We start by giving explicitly the classical breakings of the Ward
identities
(\ref{ghostequation}) and (\ref{ghostequations}) for the old formulation
or
(\ref{ghostequations2}) in the new formulation:

\eq\ba{rl}
\D^i_{\rm G}=&\intx \LP
f^{ijk}\lac -A^{*}{}^{j\m}A^k_\m +\l^{*j}\l^k
+\lb^{*j}\lb^{k} +c^{*j}c^k \rac
\es &\phantom{\intx}\
+T^i_{AB} \lac \f^{*}_A\f_B +\fb^{*}_A\fb_B
-\p^*_A\p_B -\pb^*_A\pb_B\rac
\RP\ ,\es
\O_{{\rm D}\m}=&\dsum{\vf\in I}(-1)^{{\rm gh}(\vf^*)}
                     \vf^* \pa_\m\vf
                     +\pa_\n\lp A^{*i\n}A^i_\m \rp\ ,\es
\O^{ab}_{{\rm L}}=&
-\frac{i}{2}\l^{*i}\s^{ab}\l^i
-\frac{i}{2}\lb^{*i}\s^{ab}\lb^i
-\frac{i}{2}\p_A^{*}\s^{ab}\p_A
-\frac{i}{2}\pb_A^{*}\s^{ab}\pb_A
\ ,\es
\O_{\rm R}=&\dsum{\vf\in I}(-1)^{{\rm gh}(\vf^*)}
                 \vf^* iR_\vf\vf\ .
\ea\eqn{breakings}

For the  supplementary
ghost identities of the new formulation (\ref{ghostequations3}),
the classical breakings are:

\eq\ba{rl}
\O_{\chi}=&-2\f_A\p^*_A\ ,\es
{\bar \O}_{\chib}=&-2\fb_A\pb^*_A\ ,\es
\O_{\k}=&\dsum{\vf\in I}(-1)^{{\rm gh}(\vf^*)}
                     \vf^* {\widetilde d}_{\vf}\vf\ .
\ea\eqn{breakings2}

We can then write down the nontrivial commutation rules
for the algebra formed by the ST operator
(\ref{operateurdeST})
and the operators defining the Ward identities
(\ref{gaugecondition}-\ref{ghostequations}) or
(\ref{ghostequations2}-\ref{ghostequations3}), valid for any functional
$\g$ with zero GP:

\eq
\BB_\g S(\g) =0\ ,
\eqn{ss}
\eq
\dfud{}{b^i}S(\g)
-\BB_\g\lp\dfud{\g}{b^i}-\O^i_b\rp
={\bar F}^i\g
+b\O^i_b\ ,
\eqn{sg}
%\eq
%{\bar F}^iS(\g)
%+\BB_\g {\bar F}^i\g
%=0\ ,
%\eqn{santif}
\eq
F^i S(\g)
+\BB_\g\lp F^i\g -\D^i_{\rm G}\rp
=W^i_{\rm G}\g\ ,
\eqn{sfant}
%\eq
%W^i_{\rm G}S(\g)
%-\BB_\g W^i_{\rm G}\g
%=0\ ,
%\eqn{srig}
\eq
\dfud{}{\xi^\m}S(\g)
+\BB_\g\lp\dfud{\g}{\xi^\m}-\O_{{\rm D}\m}\rp
=\om_{{\rm D}\m}\g
+\pa_\n\lp \xi^\n\dfud{\g}{\xi^\m}\rp\ ,
\eqn{sdiff}
%\eq
%\om_{{\rm D}\m}S(\g)
%-\BB_\g\lp\om_{{\rm D}\m}\g\rp
%=0 ???\ ,
%\eqn{ssdiff}
\eq
\dfud{}{\l_{ab}}S(\g)
+\BB_\g\lp\dfud{\g}{\l_{ab}}-\O_{\rm L}^{ab}\rp
=\om_{\rm L}^{ab}\g
+\pa_\m\lp \xi^\m\dfud{\g}{\l_{ab}}\rp\ ,
\eqn{slorentz}
%\eq
%\om_{\rm L}^{ab}S(\g)
%-\BB_\g\lp\om_{\rm L}^{ab}\g\rp
%=0 ???\ ,
%\eqn{sslorentz}
\eq
\dfud{}{\eta}S(\g)
+\BB_\g\lp\dfud{\g}{\eta}-\O_{\rm R}\rp
=\om_{\rm R}\g
+\pa_\m\lp \xi^\m\dfud{\g}{\eta}\rp\ .
\eqn{sr}
%\eq
%\om_{\rm R}S(\g)
%-\BB_\g\lp\om_{\rm R}\g\rp
%=0 ???\ ,
%\eqn{ssr}

Moreover, for the old formulation:

\eq
\dfud{}{w_\m}S(\g)
+\BB_\g\lp\dfud{\g}{w_\m}\rp
=\om_{\rm C}^\m\g
+\pa_\n\lp \xi^\n\dfud{\g}{w_\m}\rp
-\pa_\n\xi^\m\dfud{\g}{w_\n}
-2i E^\m\dfud{\g}{w}
\ ,
\eqn{sc}
%\eq
%\om_{\rm C}^\m S(\g)
%+\BB_\g\lp\om_{\rm C}^\m\g\rp
%=0 ???\ ,
%\eqn{ssc}
\eq
\dfud{}{w}S(\g)
-\BB_\g\lp\dfud{\g}{w}\rp
=-\pa_\m\dfud{\g}{w_\m}-\pa_\m\lp\xi^\m\dfud{\g}{w}\rp
\ ,
\eqn{sc2}

 whereas, for the new formulation:

\eq
\dfud{}{K_{\m\n}}S(\g)
+\BB_\g\lp\dfud{\g}{K_{\m\n}}\rp
=\om_K^{\m\n}\g
+ \pa_\r \lp\xi^\r\dfud{\g}{K_{\m\n}}\rp
-\pa_\r\xi^\m\dfud{\g}{K_{\r\n}}
+\pa_\r\xi^\n\dfud{\g}{K_{\r\m}}
+2\k\dfud{\g}{K_{\m\n}}
\ ,
\eqn{sk}
\eq
\dfud{}{\chi}S(\g)
-\BB_\g\lp\dfud{\g}{\chi}-\O_\chi\rp
= \om_\chi\g
- \pa_\m \lp\xi^\m\dfud{\g}{\chi}\rp
+\lp
\frac{i}{4}\l_{ab}\s^{ab}
+i\eta
+\frac{1}{2}\k
-2\lp\gP_\m\s^\m\eb\rp
\rp\dfud{\g}{\chi}
\ ,
\eqn{schi2}
%\eq
%\dfud{}{\chib}S(\g)
%-\BB_\g\lp\dfud{\g}{\chib}-{\bar \O}_\chib\rp
%= {\bar\om}_\chib\g
%- \pa_\m \lp\xi^\m\dfud{\g}{\chib}\rp
%+\lp
%\frac{i}{4}\l_{ab}\sbar^{ab}
%-i\eta
%+\frac{1}{2}\k
%-2\lp\e\s^\m\gPb_\m\rp
%\rp\dfud{\g}{\chib}
%\ ,
%\eqn{schib}
\eq
\HH_\k S(\g)
+\BB_\g\lp \HH_\k\g-\O_\k\rp
=\om_\k\g
+\pa_\m \lp\xi^\m \ \HH_\k \g\rp
\ .
\eqn{skappa}

 This algebra implies the following Ward operators:

{\bf The antighost operator}

For the old formulation:

\eq
{\bar F}^i =\displaystyle{
\dfud{}{\cb^i}
+\pa_\m\lp \sqrt{g}g^{\m\n}\dfud{}{A^{*i\n}}\rp
+2\k\dfud{}{b^i}
-\pa_\m\lp\xi^\m\dfud{}{b^i}\rp \ ,}
\eqn{oldantighost}

and for the new formulation:

\eq
{\bar F}^i =\displaystyle{
\dfud{}{\cb^i}
+\lp\pa_\m-2I_\m\rp\lp \sqrt{g}g^{\m\n}\dfud{}{A^{*i\n}}\rp
+2\k\dfud{}{b^i}
-\pa_\m\lp\xi^\m\dfud{}{b^i}\rp \ .}
\eqn{newantighost}

{\bf Ward operator for the rigid $\GG$-transformations}

\eq\ba{rl}
W_{\rm G}^i =\intx\LP &\!\!\!
-f^{jik}\lac    A^k_\m\dfud{}{A^j_\m}
                +A^{*k\m}\dfud{}{A^{*j\m}}
                +\l^k\dfud{}{\l^j}
                +\l^{*k}\dfud{}{\l^{*j}}
\right.\es &\!\!\!
\phantom{-f^{ijk}}\ \left.
                -\lb^{k}\dfud{}{\lb^{j}}
                -\lb^{*k}\dfud{}{\lb^{*j}}
                +c^k\dfud{}{c^j}
                +c^{*k}\dfud{}{c^{*j}}
                +\cb^k\dfud{}{\cb^j}
                +b^k\dfud{}{b^j}
\rac\es &\!\!\!
                -T^i_{AB}\lac
                \f_B\dfud{}{\f_A}
                +\f^{*}{}_B\dfud{}{\f^{*}{}_A}
                +\fb_B\dfud{}{\fb_A}
                +\fb^{*}{}_B\dfud{}{\fb^{*}{}_A}
                +\p_B\dfud{}{\p_A}
                +\p^{*}_B\dfud{}{\p^{*}_A}
\right.\es &\!\!\!
\phantom{-T^i_{AB}}\ \left.
                -\pb_{B}\fud{}{\pb_{A}}
                -\pb^{*}{}_{B}\fud{}{\pb^{*}{}_{A}}\rac\RP\ .
\ea\eqn{wrig}

{\bf Ward operator for the diffeomorphisms}

\eq
\om_{{\rm D}\m} = \om_{{\rm T}\m}+\pa_\n \om_{\m}{}^\n\ ,
\eqn{omegadiff1}

where

\eq
\om_{{\rm T}\m} =
\LP\sumall \pa_\m\vf\dfud{}{\vf}\RP\ ,
\eqn{omegadiff1b}

and, for the old formulation:

\eq\ba{ll}
 \om_{\m}{}^\n =&
\LP\d_\m^\n\displaystyle{\sum_{\vf^*\in I^*}\vf^*\dfud{}{\vf^*}}
-A^i_\m\dfud{}{A^i_\n}
+A^{*i\n}\dfud{}{A^{*i\m}}%\es&
-\vb{\m}{a}\dfud{}{\vb{\n}{a}}\es&
-\gP_\m\dfud{}{\gP_\n}
+\gPb_\m\dfud{}{\gPb_\n}
-C_{\m\r}\dfud{}{C_{\n\r}}
-B_\m\dfud{}{B_\n}
-w_\m\dfud{}{w_\n}
\RP\ ,
\ea\eqn{omegadiff2o}

whereas for the new formulation:

\eq\ba{ll}
 \om_{\m}{}^\n =&
\LP\d_\m^\n\displaystyle{\sum_{\vf^*\in I^*}\vf^*\dfud{}{\vf^*}}
-A^i_\m\dfud{}{A^i_\n}
+A^{*i\n}\dfud{}{A^{*i\m}}%\es&
-\vb{\m}{a}\dfud{}{\vb{\n}{a}}\es&
-\gP_\m\dfud{}{\gP_\n}
+\gPb_\m\dfud{}{\gPb_\n}
-C_{\m\r}\dfud{}{C_{\n\r}}
-B_\m\dfud{}{B_\n}
-K_{\m\r}\dfud{}{K_{\n\r}}
-I_\m\dfud{}{I_\n}
-J_\m\dfud{}{J_\n}
\RP\ .
\ea\eqn{omegadiff2n}

{\bf Ward operator for the Lorentz transformations}

For the old formulation:

\eq\ba{rl}
\om_{\rm L}^{ab} =&\lp
\frac{i}{2}\l^i\s^{ab}\dfud{}{\l^i}
-\frac{i}{2}\lb^i\s^{ab}\dfud{}{\lb^i}
+\frac{i}{2}\p_A\s^{ab}\dfud{}{\p_A}
-\frac{i}{2}\pb_A\s^{ab}\dfud{}{\pb_A}
\right.\es &
+\lp\eta^{ac}\vb{\m}{b}-\eta^{bc}\vb{\m}{a}\rp\dfud{}{\vb{\m}{c}}
+\frac{i}{2}\gP_\m\s^{ab}\dfud{}{\gP_\m}
-\frac{i}{2}\gPb_\m\s^{ab}\dfud{}{\gPb_\m}
\es &\left.
+\frac{i}{2}\e\s^{ab}\dfud{}{\e}
-\frac{i}{2}\eb\s^{ab}\dfud{}{\eb}
+2          \lp\d^a_c\l^b{}_d
              -\d^b_c\l^a{}_d\rp\dfud{}{\l_{cd}}
\rp\ ,
\ea\eqn{omegalorentz2}

and for the new formulation:

\eq\ba{rl}
\om_{\rm L}^{ab} =&\lp
\frac{i}{2}\l^i\s^{ab}\dfud{}{\l^i}
-\frac{i}{2}\lb^i\s^{ab}\dfud{}{\lb^i}
+\frac{i}{2}\p_A\s^{ab}\dfud{}{\p_A}
-\frac{i}{2}\pb_A\s^{ab}\dfud{}{\pb_A}
\right.\es &
+\lp\eta^{ac}\vb{\m}{b}-\eta^{bc}\vb{\m}{a}\rp\dfud{}{\vb{\m}{c}}
+\frac{i}{2}\gP_\m\s^{ab}\dfud{}{\gP_\m}
-\frac{i}{2}\gPb_\m\s^{ab}\dfud{}{\gPb_\m}
\es &
+\frac{i}{2}\e\s^{ab}\dfud{}{\e}
-\frac{i}{2}\eb\s^{ab}\dfud{}{\eb}
+2          \lp\d^a_c\l^b{}_d
              -\d^b_c\l^a{}_d\rp\dfud{}{\l_{cd}}
\es &\left.
+\frac{i}{2}\chi\s^{ab}\dfud{}{\chi}
-\frac{i}{2}\chib\s^{ab}\dfud{}{\chib}
\rp\ .
\ea\eqn{omegalorentz}

{\bf Ward operator for the $R$-transformations}

\eq
\om_{\rm R}=\LP
-\pa_\m\dfud{}{B_\m}
+\sumall iR_\vf\vf\dfud{}{\vf}\RP\ .
\eqn{omegar}

{\bf Ward operators for the $C$-transformations} (old formulation only)

\eq
\om_{\rm C}^\m=\LP
\pa_\n\dfud{}{C_{\m\n}}\RP\ .
\eqn{omegac}

{\bf Ward operator for the shift transformations} (new formulation only)

\eq
\om_K^{\m\n}
=\lp
\dfud{}{C_{\m\n}}
+\frac{1}{4}\pa_\l\lac
\e_\r {}^{\l\m\n}\lp
3\dfud{}{B_\r}
+\eb\sbar^\r\dfud{}{\chi}
+\e\s^\r\dfud{}{\chib}
\rp\rac\rp \ .
\eqn{omegashift}

{\bf Ward operator for the supertrace transformations} (new formulation
only)

\eq\ba{ll}
\om_\chi &
=\lp
2\f_A\dfud{}{\p_A}
-2\p^*_A\dfud{}{\f^*_A}
-\s_\m\dfud{}{\gPb_\m}
+3\s^\n\sbar_\m\gP_\n\dfud{}{B_\m}
-i\s_{ab}\e\dfud{}{\l_{ab}}\right.\es &\left.
-4\e\cb^i\dfud{}{b^i}
+3i\e\dfud{}{\eta}
+2\e\dfud{}{\k}
+2\lp\e\gP_\m\rp\s^\m\dfud{}{\chib}
+\lp 4i\s_{\m\n}\e+4\e C_{\m\n}\rp\dfud{}{K_{\m\n}}\rp\ .
\ea\eqn{omegasupertrace}

{\bf Ward operator for the (modified) Weyl transformations}
(new formulation only)

\eq\ba{lr}
\om_\k &=
\LP
\sumall {\widetilde d}_\vf\vf\dfud{}{\vf}
-\frac{1}{4}\pa_\l\lac
\e_\r {}^{\l\m\n}C_{\m\n}\lp
3\dfud{}{B_\r}
+\eb\sbar^\r\dfud{}{\chi}
+\e\s^\r\dfud{}{\chib}
\rp\rac \es &
+\pa_\m\lac
\frac{i}{2}\eb\sbar^\m\dfud{}{\chi}
-\frac{i}{2}\e\s^\m\dfud{}{\chib}
+\dfud{}{I_\m}
\rac
\RP\ .
\ea\eqn{omegaweyl}

{\bf Linear Symmetries}

Finally, setting $\g=\S$ in (\ref{sg}) to (\ref{skappa}) leads to:

\eq\ba{lll}
{\bar F}^i\S = -b\O^i_b = \O^i_\cb\ ,\hspace{1cm}&
W^i_{\rm G}\S = 0\ ,\hspace{1cm}&\es
\om_{{\rm D}\m}\S = -\pa_\n\lp\xi^\n\O_{{\rm D}\m}\rp\ ,\hspace{1cm}&
\om_{\rm L}^{ab}\S =  -\pa_\m\lp\xi^\m\O_{\rm L}^{ab}\rp\ ,\hspace{1cm}&
\om_{\rm R}\S = -\pa_\m\lp\xi^\m\O_{\rm R}\rp\ ,
\ea\eqn{symmlin}

and, for the old formulation:

\eq
\om_{\rm C}^\m\S = 0\ ,
\eqn{symlinold}

whereas for the new formulation:

\eq
\om_K^{\m\n}\S=0\ ,\hspace{1cm}
\om_\chi\S = \pa_\m\lp\xi^\m\O_\chi\rp
-\lp\frac{i}{4}\l_{ab}\s^{ab}
    +i\eta
    +\frac{1}{2}\k
    -2\lp\gP_\m\s^\m\eb\rp
\rp\O_\chi\ ,\hspace{1cm}
\om_\k\S = -\pa_\m\lp\xi^\m\O_\k\rp\ .
\eqn{symlinnew}

These are the (linearly) broken classical symmetries of the model
encoded
in the ST identity.

{\bf Local Supersymmetry}

There is no such simple Ward identity for local supersymmetry.
This is due to the nonlinearity  of the supersymmetry
transformations and therefore to the absence of a  ghost identity for the
supersymmetry ghost $\e$. We have indeed seen that
the Ward identities corresponding to the linear symmetries, a priori
contained in the Slavnov-Taylor identity, are extracted therefrom
through the action of the ghost identity operators, as explained
at the beginning of this Appendix. %\ref{appendixd}
It is however possible to write down
a Ward identity for the local supersymmetry in the F.L.
-- hence for the conservation of the spinor
current $Q^\m{}_\a$ --
which one may
define~\cite{I} as the derivative of the
Slavnov-Taylor identity with respect to $\e(x)$, setting then
$\e(x)$ as well as the remaining external ghosts to zero:

\eq\ba{ll}
\om_{{\rm S}\a} \S
 &=\left. \dfud{}{\e^\a} \SS(\S) \right|_{{\rm ext.\ ghosts}=0}\es
 &=\left.\lp \S,\dfud{\S}{\e^\a} \rp\right|_{{\rm ext.\ ghosts}=0}
   + i\pa_\m \dfud{\left.\S\right|_{{\rm ext.\ ghosts}=0}}{\Psi_\m^\a}\es
 &\    + \mbox{ terms vanishing in the flat limit} = 0\ .
\ea\eqn{susywi}

{\bf Rigid Ward Operators}

By integrating these local Ward identities, we find the
(non-broken) global Ward identities

\eq\ba{lll}
W_{{\rm D}\m}\S = 0\ ,\hspace{1cm}&
W_{\rm L}^{ab}\S =  0\ ,\hspace{1cm}&
W_{\rm R}\S = 0\ ,\es
W_K^{\m\n}\S=0\ ,\hspace{1cm}&
W_\k\S =0\ ,&
\ea\eqn{symmlinglobales}

corresponding to invariance under the translations,
the (global) Lorentz transformations,
the (global) $R$-transformations,
the (global) shift transformations and
the (global) Weyl transformations, respectively.

\section{Computation of $\S_c$}
\label{contretermes}

In this  Appendix, we solve  the cohomological problem set by
(\ref{contrbrs}),

\eq
\BB_\S\S_c=0\ ,
\eqn{contrbrs2}

in the space of local functionals in the fields, with the set of
supplementary conditions (\ref{contrgheq})

\eq
\dfud{\S_c}{b^i}=0\ ,\hspace{1cm}
F^i\S_c =0\ ,
\eqn{contrgheq2.1}
\eq
\dfud{\S_c}{\xi^\m}=\, \dfud{\S_c}{\l_{ab}}=\,\dfud{\S_c}{\eta}=0\ ,
\eqn{contrgheq2.2}
\eq
\dfud{\S_c}{K_{\m\n}}=\, \dfud{\S_c}{\chi}=\,\dfud{\S_c}{\chib}=0\ ,
\eqn{contrgheq2.3}
\eq
h_\k\S_c =0\ .
\eqn{contrgheq2.4}

The constraints (\ref{contrgheq2.1}-\ref{contrgheq2.4})
are easily solved and lead to a $\S_c$
which is independent of $b^i,\xi^\m,\l_{ab},\eta,K_{\m\n},\chi,\chib$
and which depends on $c^i$ only through its derivative $\pa_\m c^i$
and on $\k$ only through the combination

\eq
{\hat J}_\m = J_\m +\pa_\m \k\ .
\eqn{defhatj2}

We can obtain further constraints by using
(\ref{contrgheq2.1}-\ref{contrgheq2.4})
and the algebra
given in  \ref{appendixd}. For example,

\eq
0= \dfud{}{b^i} \BB_\S\S_c - \BB_\S\dfud{}{b^i}\S_c
 = {\bar F}^i\S_c\ .
\eqn{nouvellecontrainte}

In the same way, we obtain:

\eq
W^i_{\rm G}\S_c = 0\ ,
\eqn{autrescontr1}
\eq
\om_{{\rm D}\m}\S_c = \,
\om_{\rm L}^{ab}\S_c =  \,
\om_{\rm R}\S_c = 0\ ,
\eqn{autrescontr2}
\eq
\om_K^{\m\n}\S_c= 0\ ,
\eqn{autrescontr3}
\eq
\om_\k\S_c = 0\ .
\eqn{autrescontr4}
\eq
\om_\chi\S_c = \,
\om_\chib\S_c =0\ .
\eqn{autrescontr5}

(\ref{nouvellecontrainte}) and (\ref{autrescontr3}) are easily
solved and lead to a $\S_c$ that depends on $\cb^i$ and on $C_{\m\n}$ only
through the combinations

\eq\ba{ll}
{\hat A}^{*i\m} &=A^{*i\m}  +\sqrt{g} g^{\m\n}(\pa_\n +2I_\n)\cb^i\es
{\hat B}_\m &= B_\m +\frac{3}{8}\e_\m {}^{\n\r\l}\pa_\n C_{\r\l}\ .
\ea\eqn{hataethatb}

In the new variables ${\hat J}_\m ,
{\hat A}^{*i\m}$ and ${\hat B}_\m$, (\ref{contrgheq2.4}),
(\ref{nouvellecontrainte}) and (\ref{autrescontr3}) look much simpler:

\eq
\dfud{\S_c}{\k}=0
\ ,
\eqn{contrnouvvar1}
\eq
\dfud{\S_c}{\cb^i}=0
\ ,
\eqn{contrnouvvar2}
\eq
\dfud{\S_c}{C_{\m\n}}=0
\ ,
\eqn{contrnouvvar3}

and therefore we assume from now on to work with these  new
variables.

The general solution of  (\ref{contrbrs2}) can be written formally as

\eq
\S_c = \S_{\rm ph} +\BB_\S {\hat \S}_c\ .
\eqn{contrt2}

We will proceed in two steps: the first to  explicitly construct
${\hat \S}_c$ by enumerating the candidates and applying the constraints
and the second to find the cohomology element $\S_{\rm ph}$ by
filtration methods~\cite{pi-so}.

\subsection{Construction of ${\hat \S}_c$}
\label{appendixg1}

We look for the most general functional ${\hat \S}_c$ with
dimension 4, ghost number -1 and $R$-weight 0 such that $\BB_\S{\hat \S}_c$
satisfies (\ref{contrgheq2.1})-(\ref{contrgheq2.4}) and
(\ref{autrescontr1})-(\ref{autrescontr5}).

{\bf ${\hat \S}_c$ is independent of $\cb^i$ and $b^i$}:
Let us introduce the counting operator

\eq
N=\intx \lac b^i\dfud{}{b^i}+\cb^i\dfud{}{\cb^i}\rac\ .
\eqn{defn}

On one hand we have

\eq
N\lp\BB_\S{\hat \S}_c\rp=0\ ,
\eqn{contrn}

which follows form (\ref{contrgheq2.1}) and (\ref{contrnouvvar2}).
On the other hand,

\eq
\left[ N , \BB_\S \right] =0\ .
\eqn{commbsn}

So, expanding ${\hat \S}_c$ according to $N$, namely  writing

\eq
{\hat \S}_c = \dsum{n\geq 0} {\hat \S}_c^{(n)}
\hspace{1cm}\mbox{with}\hspace{6mm}
N {\hat \S}_c^{(n)} = n {\hat \S}_c^{(n)}\ ,
\eqn{expcontr}

we have

\eq
0= \left[ N , \BB_\S \right] {\hat \S}_c
 = - \BB_\S\lp N  {\hat \S}_c \rp
 = -\dsum{n\geq 0}n\ \BB_\S{\hat \S}_c^{(n)}
\ ,
\eqn{expcontr2}

which means that only ${\hat \S}_c^{(0)}$, \ie the $b^i , \cb^i$
independent part of ${\hat \S}_c$, contributes to $\BB_\S{\hat \S}_c$.
Thus, without loss of generality, we can choose ${\hat \S}_c$
independent of $b^i$ and $\cb^i$.

{\bf ${\hat \S}_c$ is independent of $C_{\m\n}$}:
As $\BB_\S{\hat \S}_c$ is constrained to be independent
of $K_{\m\n}$ (see (\ref{contrgheq2.3})), we have

\eq \ba{ll}
0&
=\dfud{}{K_{\m\n}}\lp \BB_\S{\hat \S}_c\rp
=\lac \dfud{}{K_{\m\n}} ,\BB_\S\rac {\hat \S}_c
 - \BB_\S \dfud{{\hat \S}_c}{K_{\m\n}}
\es &
=\om_K^{\m\n}{\hat \S}_c
+ \pa_\r \lp\xi^\r\dfud{{\hat \S}_c}{K_{\m\n}}\rp
-\pa_\r\xi^\m\dfud{{\hat \S}_c}{K_{\r\n}}
+\pa_\r\xi^\n\dfud{{\hat \S}_c}{K_{\r\m}}
+2\k\dfud{{\hat \S}_c}{K_{\m\n}}
- \BB_\S \dfud{{\hat \S}_c}{K_{\m\n}} \ .
\ea\eqn{indepc}

It can be easily seen by writing down the most general candidate that
${\hat \S}_c$ is independent of $K_{\m\n}, \chi$ and $\chib$,
and therefore, (\ref{indepc}) leads to

\eq
\dfud{{\hat \S}_c}{C_{\m\n}}=0\ .
\eqn{indepc2}

{\bf ${\hat \S}_c$ is independent of $c^i$ and satisfies
$W^i_{\rm G}{\hat \S}_c=0$}:
We can write

\eq
{\hat \S}_c = \intx\lp
Q^{ij}(\vb{\m}{a}) c^{*i}c^j +\;\mbox{independent of}\; c^i\rp\ ,
\eqn{indepc3}

where the scalar $Q^{ij}(\vb{\m}{a})$ is a  local functional
in the vierbein. From (\ref{contrgheq2.1}), we get

\eq
\intx \dfud{}{c^i}\lp\BB_\S {\hat \S}_c\rp =0\ ,
\eqn{indepc4}

which leads, by an explicit calculation of the term linear in $c^i$, to
$Q^{ij}(\vb{\m}{a})=0$. Thus, ${\hat \S}_c $ is independent of $c^i$.
Then, we write

\eq
0
=\intx \dfud{}{c^i}\lp\BB_\S {\hat \S}_c\rp
=\lac \intx \dfud{}{c^i},\BB_\S \rac{\hat \S}_c
  - \BB_\S \intx \dfud{{\hat \S}_c}{c^i}
=W^i_{\rm G}{\hat \S}_c\ ,
\eqn{indepc5}

which shows that ${\hat \S}_c$ is invariant under
the  $\GG$-rigid transformations.

{\bf ${\hat \S}_c$ is independent of $\xi^\m$}:
We can write

\eq
{\hat \S}_c = \intx\lp
P_\m \xi^\m +\;\mbox{independent of}\; \xi^\m\rp\ ,
\eqn{indepc6}

where $P_\m$ is a functional of dimension 5, ghost number -2 and $R$-weight
0.
{}From the condition $\dfud{}{\xi^\m}\BB_\S {\hat \S}_c=0$, we get $P_\m=0$
and
thus ${\hat \S}_c$ is independent of $\xi^\m$.

{\bf Remaining candidates}:
We can now list the remaining functionals potentially
constituting ${\hat \S}_c$

\eq\ba{rl}
{\hat \S}_c = \intx \LP &\!\!\!
    \displaystyle{
Z_A  {\hat A}^{*i\m}A^i_\m
+Z_\l \l^{*i}\l^i
+{\bar Z}_\lb \lb^{*i}\lb^i
+Z_{\f AB} \f^*_A\f_B
+{\bar Z}_{\fb AB}\fb^*_A\fb_B
                  }
\\
&\!\!\!
    \displaystyle{
+Z_{\p AB}\p^*_A\p_B
+{\bar Z}_{\pb AB}\pb^*_A\pb_B
+Y_{AB}  \lp\gPb_\m\sbar^\m\p^*_A\rp\f_B
-{\bar Y}_{AB} \lp\gP_\m\s^\m\pb^*_A\rp\fb_B
                  } \vspace{3mm}
\\
&\!\!\!
    \displaystyle{
+V g^{\m\r}\lp \gP_\m +\frac{i}{3}\gP_\n \s^\n{}_\m\rp\l^{*i} A^i_\r
+W g^{\m\r}\lp \gP_\m\l^{*i}\rp A^i_\r
                  }
\\
&\!\!\!
    \displaystyle{
+{\bar V} g^{\m\r}\lp \gPb_\m +\frac{i}{3}\gPb_\n \sbar^\n{}_\m\rp
                       \lb^{*i} A^i_\r
+{\bar W} g^{\m\r}\lp \gPb_\m\lb^{*i}\rp A^i_\r
                  }
\RP\ ,
\ea\eqn{contretermesres}

where $Z_A,Z_\l,V$ and $W$ are arbitrary constants and
$Z_{\f AB},Z_{\p AB}$ and $Y_{AB}$ are $\GG$-invariant tensors.

The last step is to apply $\BB_\S$ on this ${\hat \S}_c$ and to impose the
remaining constraints, namely (\ref{contrgheq2.3}), which lead to

\eq\ba{l}
W= {\bar W} = 0\\
Z_{\f AB} +Z_{\p AB} -2Y_{AB} =0\\
{\bar Z}_{\f AB} +{\bar Z}_{\p AB} -2{\bar Y}_{AB} =0\ .
\ea\eqn{conditionsalpha}

Thus, finally, ${\hat \S}_c$ is given by

\eq\ba{rl}
{\hat \S}_c = \intx \LP &\!\!\!
    \displaystyle{
Z_A A^i_\m {\hat A}^{*i\m}
+Z_\l \l^{*i}\l^i
+{\bar Z}_\lb \lb^{*i}\lb^i
                  }
\\
&\!\!\!
    \displaystyle{
+Z_{\f AB} \lp \f^*_A +\frac{1}{2}\gPb_\m\sbar^\m\p^*_A \rp\f_B
+{\bar Z}_{\f AB}\lp \fb^*_A -\frac{1}{2}\gP_\m\s^\m\pb^*_A\rp\fb_B
                  }
\\
&\!\!\!
    \displaystyle{
+Z_{\p AB}\p^*_A\lp \p_B +\frac{1}{2}\s^\m \gPb_\m\f_B\rp
+{\bar Z}_{\p AB}\pb^*_A\lp \pb_B-\frac{1}{2}\sbar^\m \gP_\m\fb_B\rp
                  }
\\
&\!\!\!
    \displaystyle{
+V g^{\m\r}\lp \gP_\m +\frac{i}{3}\gP_\n \s^\n{}_\m\rp\l^{*i} A^i_\r
+{\bar V} g^{\m\r}\lp \gPb_\m +\frac{i}{3}\gPb_\n \sbar^\n{}_\m\rp
                       \lb^{*i} A^i_\r
                  }
\RP\ .
\ea\eqn{contretermesres2}

The interpretation of the different terms is given in the text, after
(\ref{explicites2}).

\subsection{Computation of the Cohomology Element $\S_{\rm ph}$}

Let us denote by $\FF$ the space of local functionals in the fields
with dimension 4, ghost number 0 and $R$-weight 0,
constrained by (\ref{contrgheq2.1}-\ref{contrgheq2.4}) and
(\ref{autrescontr1}-\ref{autrescontr4})\footnote{
\equ{autrescontr5} is not included in the set of constraints defining
$\FF$ since the filtration operator we are just going to introduce
\equ{filtration2} does not commute with $\om_\chi$. Nevertheless,
as $\om_\chi$ is contained in the ST linearized operator $\BB_\S$,
the cohomology elements $\S_{\rm ph}$ we are going to construct will be
invariant under $\om_\chi$ thanks to \equ{contrbrs2}. This accounts for the fact that the constraints
(\ref{autrescontr1}-\ref{autrescontr5}) are contained in the
$\BB_\S$-invariance constraint \equ{contrbrs2} and thus do not need to be
imposed separately.
}.
We are thus looking for the
cohomology of $\BB_\S$ in $\FF$,
denoted by $\cohom{\FF}{\BB_\S}$.
To do this, let us introduce the filtration operator

\eq
N=\intx\lac
\e\dfud{}{\e}-\eb\dfud{}{\eb}
+\gP_\m \dfud{}{\gP_\m}-\gPb_\m\dfud{}{\gPb_\m}
+\chi\dfud{}{\chi}-\chib\dfud{}{\chib}
\rac\ ,
\eqn{filtration2}

which leads to the following decomposition for $\BB_\S$
and $\FF$:

\eq\ba{ll}
\BB_\S &= \ddsum{k=0}{4} \BB_\S^{(k)}\ ,
\hspace{1cm}\mbox{with}\hspace{5mm}
[N,\BB_\S^{(k)}]=k\BB_\S^{(k)}\ ,
\es
\FF &= \ddsum{k=0}{8} \FF^{(k)}\ ,
\hspace{1cm}\mbox{with}\hspace{5mm}
N\Xi^{(k)}=k\Xi^{(k)}
\hspace{3mm}\mbox{if}\hspace{3mm}
\Xi^{(k)}\in\FF^{(k)}\ .
\ea\eqn{decomp}

Writing down explicitly a basis of $\FF$
shows that there is no element of order greater than 8.

The procedure we apply in order to
find  $\cohom{\FF}{\BB_\S}$ was developed
in \cite{diximp} and applied, in particular,
in \cite{I,white1,white2,magg1,magg2,mpr,brandt}. It
consists in the following steps:

{\bf Step 1}: It has been shown in \cite{diximp} that for each
element of $\cohom{\FF}{\BB_\S}$,
we can pick up a particular representative which has the
property that its lowest order according to the filtration $N$
is in the cohomology of $\BB_\S^{(0)}$.
Thus, the first step consists in constructing
the cohomology of $\BB_\S^{(0)}$ in the different spaces $\FF^{(k)}$,
denoted by $\cohom{\FF^{(k)}}{\BB_\S^{(0)}}$, for
$k=1,\cdots,8$.

{\bf Step 2}: We then  proceed to the construction of the
``extension" $\ \cohomext{\FF^{(k)}}{\BB_\S^{(0)}}\ $ of
$\ \cohom{\FF^{(k)}}{\BB_\S^{(0)}}\ $,
defined by

\eq\ba{lll}
\cohomext{\FF^{(k)}}{\BB_\S^{(0)}}
=\Lac\Xi\in\FF\; \Bv&
\Xi=\Xi^{(k)}+\dsum{n>k}{\bar \Xi}^{(n)}
&\mbox{with}\;\;
N{\bar \Xi}^{(n)}=n{\bar \Xi}^{(n)}\;\;\mbox{for}\;\;n>k
\es&&\mbox{and}\;\;\Xi^{(k)}\in\cohom{\FF^{(k)}}{\BB_\S^{(0)}}
\es&\mbox{and}\;\;
\BB_\S\Xi=0
\Rac\ .&
\ea\eqn{cohomext}

In other words, $\cohomext{\FF^{(k)}}{\BB_\S^{(0)}}$ is the subspace
of $\FF$ which contains the $\BB_\S$-invariants whose lowest order
according to the filtration $N$ are elements of
$\cohom{\FF^{(k)}}{\BB_\S^{(0)}}$.
To perform this construction, we write explicitly
the constraint of $\BB_\S$-invariance of $\Xi$:

\eq\ba{lll}
\BB_\S\Xi=0\hspace{5mm}&
\Longleftrightarrow\hspace{5mm}&\lac\ba{l}
\BB_\S^{(0)}\Xi^{(k)}=0\es
\BB_\S^{(1)}\Xi^{(k)}+\BB_\S^{(0)}{\bar \Xi}^{(k+1)}=0\es
\BB_\S^{(2)}\Xi^{(k)}+\BB_\S^{(1)}{\bar \Xi}^{(k+1)}
+\BB_\S^{(0)}{\bar \Xi}^{(k+2)}=0\es
\vdots\ea\right.
\ea\eqn{explbinv}

We readily see that the first equation is automatically satisfied since
$\ \Xi^{(k)}\in\cohom{\FF^{(k)}}{\BB_\S^{(0)}}\ $, and that the other
ones are really constraints since, for example,
$\ \BB_\S^{(1)}\Xi^{(k)}\ $ is not automatically $\BB_\S^{(0)}$-trivial.
Thus, step 2 consists in solving this system of equations (which is finite
due to the decomposition~\equ{decomp} being finite), starting with the
cohomologies found in step 1.

{\bf Step 3}: We finally test the triviality of the elements found
in step 2 by the following construction. Let us define the sum:

\eq
\cohomext{\FF}{\BB_\S}=
\bigoplus_{k=0}^{8}\cohomext{\FF^{(k)}}{\BB_\S^{(0)}}\ ,
\eqn{cohomexttot}

as well as the subspace $\invbs{\FF}{\BB_\S}$
containing the trivial $\BB_\S$-invariants:

\eq
\invbs{\FF}{\BB_\S}=
\Lac
\Theta\in\FF\;\Bv \;
\BB_\S\Theta=0\;\;\mbox{and}\;\;
\exists\, {\hat\Theta}\;\;\mbox{such that}\;\;
\BB_\S {\hat\Theta}=\Theta
\Rac\ .
\eqn{definvbrs}

\underline{Remark}: ${\hat\Theta}$ is {\em not} subject to the
constraints defining
$\FF$, \ie it is a general local functional in the fields with dimension
3, ghost number -1 and $R$-weight 0 such that $\BB_\S{\hat\Theta}$ belongs
to $\FF$.

Thus, the cohomology $\cohom{\FF}{\BB_\S}$ is given by
the quotient of $\cohomext{\FF}{\BB_\S}$ by $\invbs{\FF}{\BB_\S}$:

\eq
\cohom{\FF}{\BB_\S}=\cohomext{\FF}{\BB_\S}/\invbs{\FF}{\BB_\S}\ .
\eqn{defcohom}

Moreover, $\invbs{\FF}{\BB_\S}$ can easily be calculated since
it is simply given by the $\BB_\S$-variation of the general
element displayed in~\ref{appendixg1}, equation \equ{contretermesres2}.

We have now finished with the description of the general method and
turn to the specific case of interest
in this paper.

{\bf Step 1}:
Due to the constraints defining $\FF$,
the action of $\BB_\S^{(0)}$ on the elements of $\FF$ reduces to:

\eq\ba{rl}
\BB_\S^{(0)}=&
S_{\GG}+S^{*(0)}+
\intx\lac
+{\hat J}_\m\dfud{}{I_\m}
\right.\es &
+\lp
-i\nabla_\m\e
+\chib\sbar_\m
+\frac{1}{4}X_\m\e
-\frac{i}{8}X^\n\e\s_{\m\n}
\rp\dfud{}{\gP_\m}
\es &\left.
-\lp
-i\nabla_\m\eb
+\chi\s_\m
-\frac{1}{4}X_\m\eb
+\frac{i}{8}X^\n\eb\sbar_{\m\n}
\rp\dfud{}{\gPb_\m}
\rac
\ ,
\ea\eqn{bsigma0}

with $X_\m$ defined in \equ{dercovsupersym},

\eq\ba{rl}
S_{\GG}=\intx&\displaystyle{
\!\!\!\lac
\nabla_\m c^i\dfud{}{A^i_\m}
-\frac{1}{2}f^{ijk}c^jc^k\dfud{}{c^i}
-f^{ijk}c^j\lp
{\hat A}^{*k\m}\dfud{}{{\hat A}^{*i\m}}
+\l^k\dfud{}{\l^i}
+\l^{*k}\dfud{}{\l^{*i}}
\right.\right.}\es&\displaystyle{\left.\left.
-\lb^{k}\dfud{}{\lb^{i}}
-\lb^{*k}\dfud{}{\lb^{*i}}
+c^{*k}\dfud{}{c^{*i}}
\rp
-T^i_{AB}c^i\lp
                \f_B\dfud{}{\f_A}
                +\f^{*}{}_B\dfud{}{\f^{*}{}_A}
                +\fb_B\dfud{}{\fb_A}
\right.\right.}\es&\displaystyle{\left.\left.
                +\fb^{*}{}_B\dfud{}{\fb^{*}{}_A}
%\right.\right.}\es&\displaystyle{\left.\left.
                +\p_B\dfud{}{\p_A}
                +\p^{*}_B\dfud{}{\p^{*}_A}
                -\pb_{B}\fud{}{\pb_{A}}
                -\pb^{*}{}_{B}\fud{}{\pb^{*}{}_{A}}\rp
\rac}\ ,
\ea\eqn{sgauge}

corresponding to the $\GG$-gauge BRS operator and

\eq\ba{l}
\ba{rl}
S^*=\intx&\LAC
\LP\,\dsum{\vf\in \{A^i_\m,\l^i,\f_A,\p_A\}}\dfud{\S_{\rm inv}}{\vf}
\dfud{}{\vf^*}\RP \es
& +\lp\nabla_\m{\hat A}^{*i\m}
                     +f^{ijk}(\l^{*j}\l^k+\lb^{*j}\lb^k)
                     +T^i_{ab}(\f^*_a\f_b+\fb^*_a\fb_b
                               -\p^*_a\p_b-\pb^*_a\pb_b)
\rp\dfud{}{c^{*i}}\RAC\ ,
\ea
\es
S^*=\ddsum{k=0}{4}S^{*(k)}\ ,
\hspace{8mm}\mbox{with}\hspace{5mm}
[N,S^{*(k)}]=kS^{*(k)}\ ,
\ea\eqn{sstar}

where $\S_{\rm inv}$ is given by \equ{nouvelleaction}.

We see that $(I_\m,{\hat J}_\m)$ appear in $\BB_\S^{(0)}$
as a BRS-doublet, from what
follows\footnote{see e.g. Section 5.2. of \cite{pi-so}}
that the $\BB_\S^{(0)}$-cohomology can be choosen independent
of these fields.

The calculation of the cohomology spaces
$\cohom{\FF^{(k)}}{\BB_\S^{(0)}},\; k=0,\ldots,8\ $ is then standard:
we write the most general element of
$\FF^{(k)}$ which is independent of $(I_\m,{\hat J}_\m)$,
apply the condition of $\BB_\S^{(0)}$-invariance
and finally test the triviality of the remaing candidate.
We give here only $\cohom{\FF^{(0)}}{\BB_\S^{(0)}}$ since,
anticipating the Step 2, we found that
the extensions of $\cohom{\FF^{(k)}}{\BB_\S^{(0)}}$ for $k=2,\ldots,8$
are all empty. Thus, the most general element $\Xi^{(0)}$
of $\cohom{\FF^{(0)}}{\BB_\S^{(0)}}$ is:

\eq\ba{rl}
\Xi^{(0)} = \intx \sqrt{g}\LP &\!\!\!
    \displaystyle{
    \a_1 F^{i\m\n}F^i_{\m\n}
  +\a_{2(AB)(CD)} \f_A\f_b\fb_C\fb_D
  +\a^i_{3AB}\f_A(\pb_B\lb^i)
                  }
\\
&\!\!\!
    \displaystyle{
  +{\bar \a}^i_{3AB}\fb_A(\p_B\l^i)
  +\a_{4(AB)C}(\p_A\p_B)\f_C
  +{\bar \a}_{4(AB)C}(\pb_A\pb_B)\fb_C
                  }
\\
&\!\!\!
    \displaystyle{
  +\a_5 C^{\m\n\r\l} C_{\m\n\r\l}
  +\a_6 {\hat G}^{\m\n} {\hat G}_{\m\n}
                  }
\RP\ ,
\ea\eqn{cohom00}

where $\a_1,\a_5,\a_6$ are arbitrary constants and
$\a_2,\a_3,\a_4$ are $\GG$-invariant tensors.
The Weyl tensor $C_{\m\n\r\l}$ is

\eq
C_{\m\n\r\l}=
R_{\m\n\r\l}
-\frac{1}{2}\lp
g_{\m\r}R_{\n\l}
-g_{\n\r}R_{\m\l}
-g_{\m\l}R_{\n\r}
+g_{\n\l}R_{\m\r}
\rp
+\frac{1}{6}R\lp
g_{\m\r}g_{\n\l}
-g_{\m\l}g_{\n\r}
\rp
\ ,
\eqn{weyltensor}

where the Riemann tensor $R_{\m\n\r\l}$,
the Ricci tensor $R_{\m\n}$ and the scalar curvature $R$
are defined in \ref{appendixb}, equation (\ref{riemann}).

Finally, the $R$-curvature tensor ${\hat G}_{\m\n}$ is

\eq
{\hat G}_{\m\n}=\pa_\m {\hat B}_\n-\pa_\n {\hat B}_\m
\ .
\eqn{rcurvtensor}

{\bf Step 2 and 3}:
The extension of $\cohom{\FF^{(k)}}{\BB_\S^{(0)}}$ is then straightforward
(although going through tedious calculations) and proceeds as explained
in solving \equ{explbinv}.
Finally, showing that the extended functionals are not trivial leads to
the following most general element $\S_{\rm ph}$ of $\cohom{\FF,\BB_\S}$:

\[
\S_{\rm ph}=Z_G \dpad{\S}{G}
+Z_{(ABC)}\dpad{\S}{\L_{(ABC)}}
+{\bar Z}_{(ABC)}\dpad{\S}{\Lb_{(ABC)}}
+\S_{\rm CSG}
\ ,
\]

where $\S_{\rm CSG}$ is the purely conformal supergravity action~\cite{csg},
which is of the form

\eq
\S_{\rm CSG}=Z_{\rm CSG}\intx\sqrt{g}\lp
C^{\m\n\r\l} C_{\m\n\r\l}
-\frac{3}{4} {\hat G}^{\m\n} {\hat G}_{\m\n}
+\cdots
\rp
\ .
\eqn{csg0}

The dots involve the gravitational fields $\vb{\m}{a},
\gP_\m,{\hat B}_\m$ and contain among other a
kinetic part for the gravitino $\gP_\m$.

The interpretation of these different counterterms is done
in the text, just after~\equ{explicites2}.

%************************************************************************

\end{document}